\documentclass[a4paper,12pt]{article}
\usepackage{tikz} 
\usepackage{amssymb}
\usepackage{algorithm}
\usepackage{algorithmic}

\RequirePackage{amsthm}
\RequirePackage{enumerate}
\RequirePackage{color}
\RequirePackage{placeins}
\RequirePackage{pdfsync}
\RequirePackage{mathbbol}
\RequirePackage{natbib}
\RequirePackage{amsmath}
\RequirePackage{amsthm}
\RequirePackage{hyperref}
\RequirePackage{xcolor}
\RequirePackage{array}
\RequirePackage{booktabs}
\RequirePackage{fullpage}
\RequirePackage{graphicx} 
\RequirePackage{comment}
\RequirePackage{multirow}

\newcommand\norm[1]{\left\lVert#1\right\rVert}
\newcommand\homog[1]{\left| #1\right|}

\newtheorem{theorem}{Theorem}[section]
\newtheorem{definition}[theorem]{Definition}
\newtheorem{lem}{Lemma}

\newtheorem{prop}[theorem]{Proposition}

\newcommand{\T}{^{\mathrm{\scriptscriptstyle T}}} 

\newcommand{\E}{{\mathbb{E}}}

\renewcommand\P{\mathbb{P}}

\def\binfty{\boldsymbol \infty}

\def\bzero{\boldsymbol 0}
\def\bone{\boldsymbol 1}

\def\bC{\boldsymbol C}

\def\bJ{\boldsymbol J}
\def\bP{\boldsymbol P}

\def\Rad{\norm{\boldsymbol X}}

\def\bU{\boldsymbol U}
\def\bV{\boldsymbol V}
\def\bX{\boldsymbol X}

\def\bY{\boldsymbol Y}
\def\bZ{\boldsymbol Z}

\def\bL{\boldsymbol L}

\def\ba{\boldsymbol a}

\def\bv{\boldsymbol v}
\def\bx{\boldsymbol x}

\def\bz{\boldsymbol z}
\def\ba{\boldsymbol a}

\def\bgamma{\boldsymbol \gamma}

\title{Multivariate distributional modeling  of low, moderate, and large   intensities without threshold selection steps}
\author{Carlo Gaetan
	\\ Dipartimento di Scienze Ambientali, Informatica e Statistica,\\ Universit\`a Ca' Foscari di Venezia, Venezia, Italy
	 \\
	 {gaetan@unive.it}
	 \\
	 \,
	 \\
	Philippe Naveau 
 	\\ Laboratoire des Sciences du Climat et l'Environnement \\
 	CNRS,
 	Gif-sur-Yvette, France\\
 	 	philippe.naveau@lsce.ipsl.fr
}
\date{\today}

\begin{document}

\maketitle 
\begin{abstract}

In fields such as hydrology and climatology, modelling the entire distribution of positive  data is essential, as stakeholders require insights into the full range of values, from low to extreme. Traditional approaches often segment the distribution into separate regions, which introduces subjectivity and limits coherence. This is especially true when dealing with multivariate data.

In line with multivariate extreme value theory, this paper presents a unified, threshold-free framework for modelling marginal behaviours and dependence structures based on an extended generalized Pareto distribution (EGPD).   We propose decomposing multivariate data into radial and angular components. The radial component is modelled using a semi-parametric EGPD and the angular distribution is permitted to vary conditionally. This approach allows for sufficiently flexible dependence modelling.

The hierarchical structure of the model facilitates the inference process.  First, we combine classical maximum likelihood estimation (MLE) methods with semi-parametric approaches based on Bernstein polynomials to estimate the distribution of the radial component. Then, we use multivariate regression techniques to estimate the angular component's parameters.

The model is evaluated through synthetic simulations and applied to hydrological datasets to exemplify its capacity to capture heavy-tailed marginals and complex multivariate dependencies without threshold specification.
\end{abstract}
\section{Introduction}
\label{sec:introduction}

In hydrology and climatology, the distribution of positive intensities such as river discharges has always been of interest for various communities. 
Flood  risk managers often focus on the analysis of high river flows. 
In contrast, farmers may be interested in periods of low river runoffs to prevent food production shortages, while energy producers in charge of electrical dams can be concerned by the full range of the variable of interest. Hence, instead of making three distinct analysis for modeling separately low, moderate and high river discharge distributions, it would be convenient to work with a single probability density distribution (pdf). 
This statement holds for a specific location on the river bank, but it is also true for sites along the same river basin as nearby measurements can be strongly dependent, i.e.  a multivariate pdf should be ideal to capture marginal river flows and their dependence structure as well.

Extreme value theory (EVT) provides a solid mathematical foundation to model the extremal tail behaviors of most distributions \citep[see e.g.][]{Coles2001}. This theory has mainly been applied to  the upper right side of the distribution (large values), but it is also valid for  the left hand side  (low values).
In practice, exceedances above  a large threshold are  fitted by 
a generalized Pareto distribution (GPD) 
\citep[see e.g.][]{dehaan:ferreira:2006}.  A delicate point for practitioners is to find the appropriate  threshold, the GPD parameters estimates being sensitive to this choice. 
To  diminish the negative effects of selecting a high  threshold, 
different approaches have been proposed to extend the GPD class 
\cite[see, e.g.][]{scarrott:macdonald:2012,Murphy:Tawn:Varty:2024}.
A  strategy is to add  a few parameters to a GPD  to allow a smooth transition from moderately large extremes to very  large  ones.  
In the univariate context, \cite{PAPASTATHOPOULOS2013131} coined the term 
extended generalized Pareto distribution (EGPD) for this task \cite[see also the recent review by][]{Naveau25}.
Still, this approach  neither capture low extremes behaviors  nor  the bulk of the distribution. As a  technique aims at  modeling  moderate and high extremes, a threshold is still needed to define moderate extremes, although this choice is less sensitive.

To completely  bypass this tricky step of selecting thresholds, \cite{Naveau2016} and \cite{Tencaliec:Favre:Naveau:Prieur:Nicolet:2020} proposed a semi-parametric model in compliance with EVT for both low and large extremes and allowing a smooth transition towards the distribution bulk. 
In this context,  this new class was also called the EGPD and this particular sense   will be kept in this paper. 
In terms of geophysical applications,  rainfall amounts were adequately fitted by a EGPD  in many case studies, see e.g. \cite{evin2018} and \cite{Rivoire22}.  
Clustering rainfall data into different regions was implemented by \citet{Legall2022} with an EGPD metric based.  
\citet{Haruna23}
also studied intensity-duration-frequency curves. Besides rainfall data, the EGPD was also used for other atmospheric variables. For example, 
\cite{turkman2021} built on the EGPD to propose and study a spatio-temporal hierarchical model for calibration purposes of wind data.  
\cite{Gamet22}  fitted an EGPD to  
daily summer temperature maxima recorded in Québec.
Incorporating covariates via EGPD parameters can also be implemented in a Bayesian context \citep{Carvalho:etal:2022}. 
As the EGPD appears to have brought flexibility to practitioners interested in modeling the full distribution of univariate variable of interest, low and high extremes included,  the extension of the EGPD into a multivariate framework should  be certainly welcome.

A first key aspect to fulfill this objective is to precisely define  low and high extremes in a multivariate context. A second issue is to find a model that  smoothly combines the multivariate bulk distribution with the modeling of the upper and lower parts. 
For the first point, 
multivariate EVT offers now  well developed mathematical  models with various  inferential schemes exist \citep[see, e.g.][]{beirlant:goegebeur:teugels:segers:2004, dehaan:ferreira:2006,DavisonHuser15}. Such existing strategies to define and model in multivariate  upper tails will be leveraged.
As for the univariate EGPD, our main requirement is to avoid introducing thresholds.
This is particularly true in a multivariate context as the threshold selection issue increases with the dimension. 
In addition, our model should be in compliance with the multivariate EVT for both high and low extremes.   
Finally, each marginal should be represented in a univariate EGPD. 
As the main field of applications we have in mind is hydrology, we focus on multivariate random variables that are non-negative and each marginal has heavy tail distributions. 
This particular setup simplifies some of the aspects, but the main ideas developed here could be extended, as is pointed out in the concluding section of this paper.

As for the univariate case, there have been   various proposals to create models that smoothly  transition from moderate to large multivariate extremes.   
In the bivariate context, one of the first examples  was proposed by\citet{VracNaveauBrobinsky07} who  modeled the entire range of positive precipitation   with   a bivariate mixture model those weights change according to  a smooth dynamic weighting function. 
Later on, \citet{Leonelli:Gamerman:2020} employed an extreme value mixture model combined with a mixture of copulas to estimate both the marginal distributions and the dependence structures.
\citet{Hu:Swallow:Castro-Camilo:2024}  proposed  a multivariate mixture model that assumes a parametric model to capture the bulk of the distribution, which is in the max-domain of attraction  of a multivariate extreme value distribution. 
The tail is described by the multivariate generalized Pareto distribution \citep{Rootzen:Tajvidi:2006}. Still, the specification of such  models requires fixing  thresholds.
\citet{andre2024joint} suggested a model that combines two bivariate copulas using a mixture. One copula is used to model dependence in the bulk, and the other is used to model dependence in the tail. This piecing-together approach was originally proposed by \citet{Aulbach:Bayer:Falk:2012} and \citet{Aulbach:Falk:Hofmann:2012}.
The weights of the mixture depend on the region (bulk or tail) that is being modeled, allowing for a smooth transition between them.
In this respect, the subjective determination of thresholds is bypassed in the copula, but it remains when we want to model the marginal distributions \citep[see][Section 4.1]{andre2024joint}. 
Also all the aforementioned articles did not study the modeling of low extremes. In this work, one important  objective is to insure that MEVT can be applied on both sides of our data range.  To reach this goal, 
we build on the classical MEVT transformation  that decomposes the data at hand into a  radial component and an angular component. 
As in  
\citet{Wadsworth:Tawn:Davison:Elton:2017}, 
MEVT points towards the assumption of independence between the angular and radial variables at large radii. Classically,  the upper tail of the radial variable is approximated  by  a GPD. In contrast, 
our radial component will be  modeled using an EGPD. In addition,  we allow that the semi-parametric distribution for the angular variables  can  depend on the radial variable, the degree of dependence will vary conditionally to the radial value.  This modeling approach also bears some similarities to recent proposals in the literature, such as those by 
\citet{murphy2024inference} and \citet{Mackay:Jonathan:2024}.  In these works, the angular variables were first modeled, followed by the modeling of the distribution of the radius given these angular variables.

The structure of the paper is as follows. In Section \ref{sec:megpd}, the main features of the multivariate EGPD are presented, with particular attention to definitions and properties, along with a semi-parametric model for the bivariate case. In Section \ref{sec:inference}, a two-step inferential procedure for the bivariate case is proposed.
Section \ref{sec:numerical_examples} introduces synthetic examples of different types of dependence and fits the semi-parametric formulation to real data from a motivating example. Finally,   Section \ref{sec:conclusions} presents the paper's conclusions.

\section{Extended generalized Pareto models}\label{sec:megpd}

\subsection{Univariate extended generalized Pareto distribution}

\begin{figure}[!ht]
	\begin{center}
		\begin{tabular}{cc}
			\includegraphics[scale=.43]{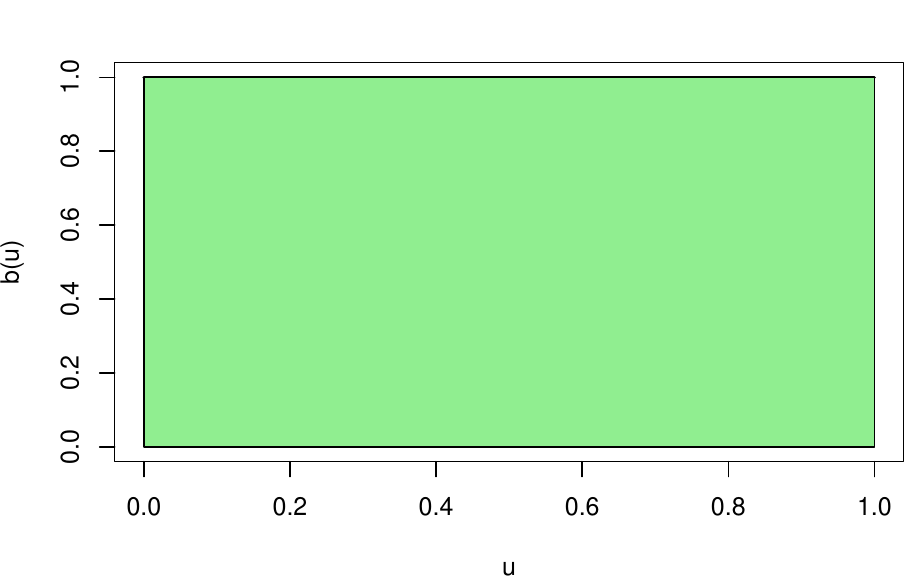} &
			\includegraphics[scale=.43]{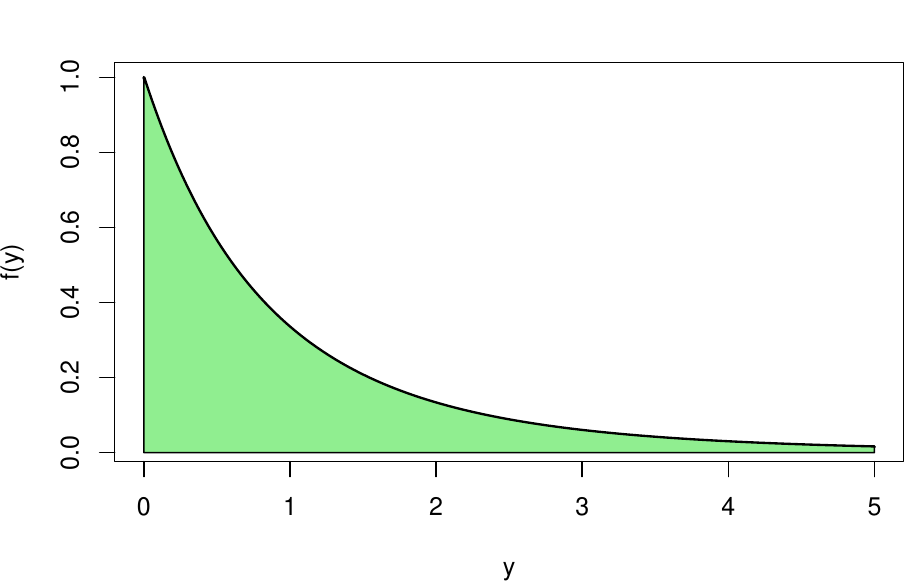}    \\
			\includegraphics[scale=.43]{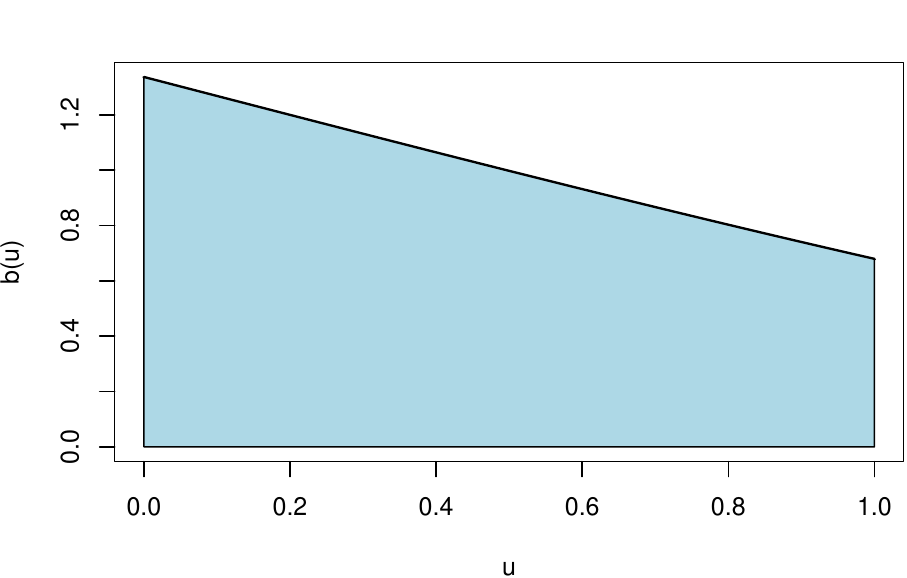} &
			\includegraphics[scale=.43]{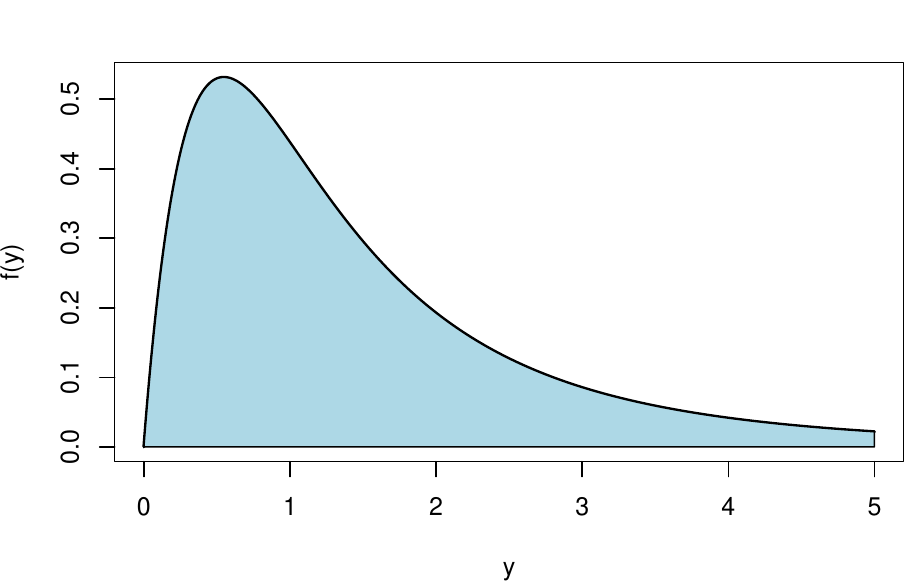}    \\
			\includegraphics[scale=.43]{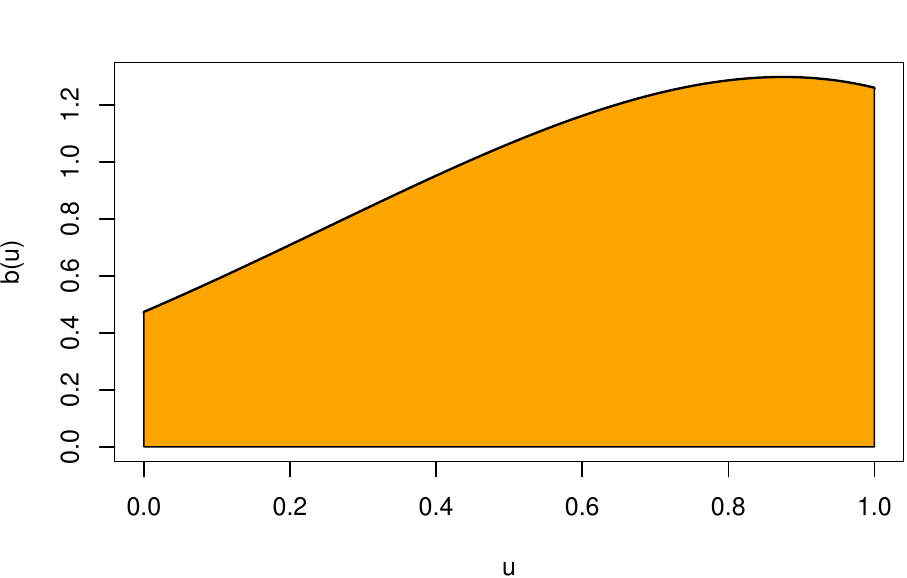} &
			\includegraphics[scale=.43]{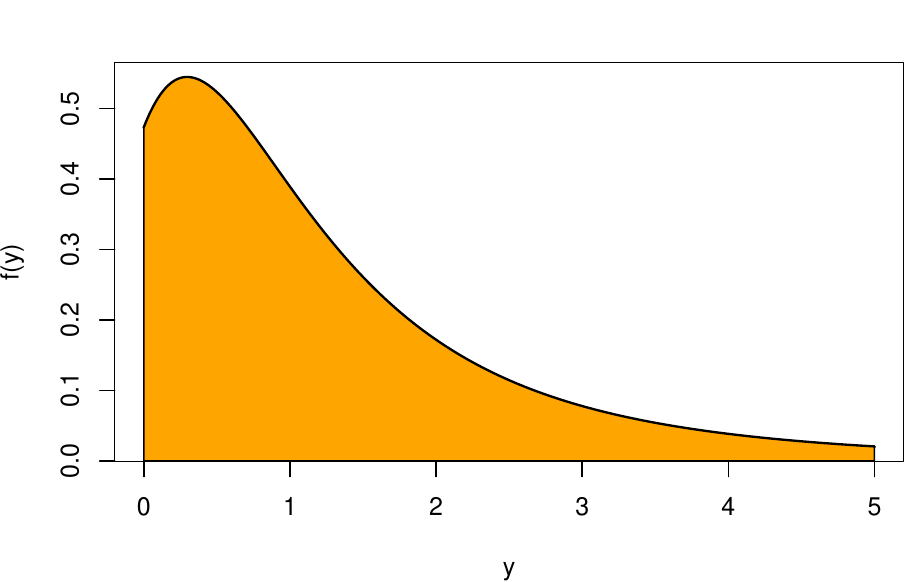}    \\
			\includegraphics[scale=.43]{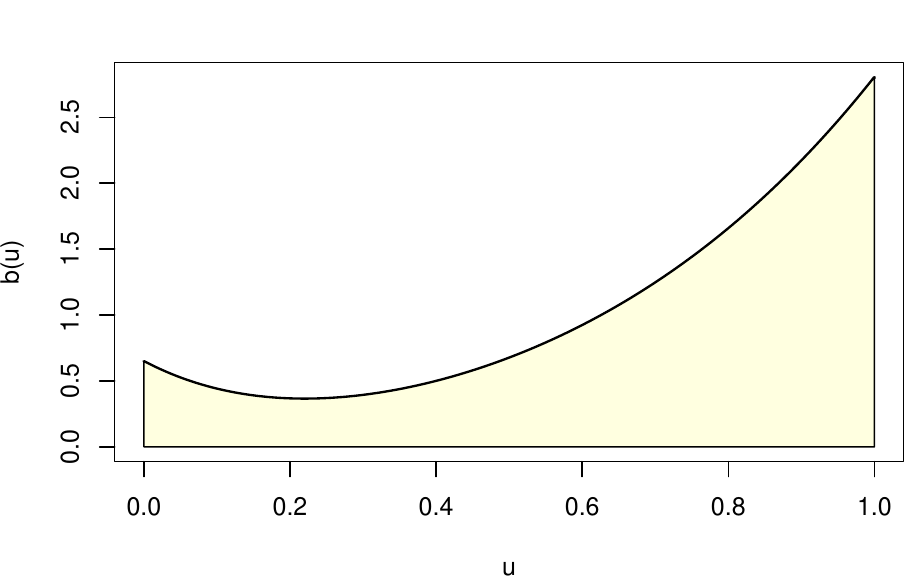} &
			\includegraphics[scale=.43]{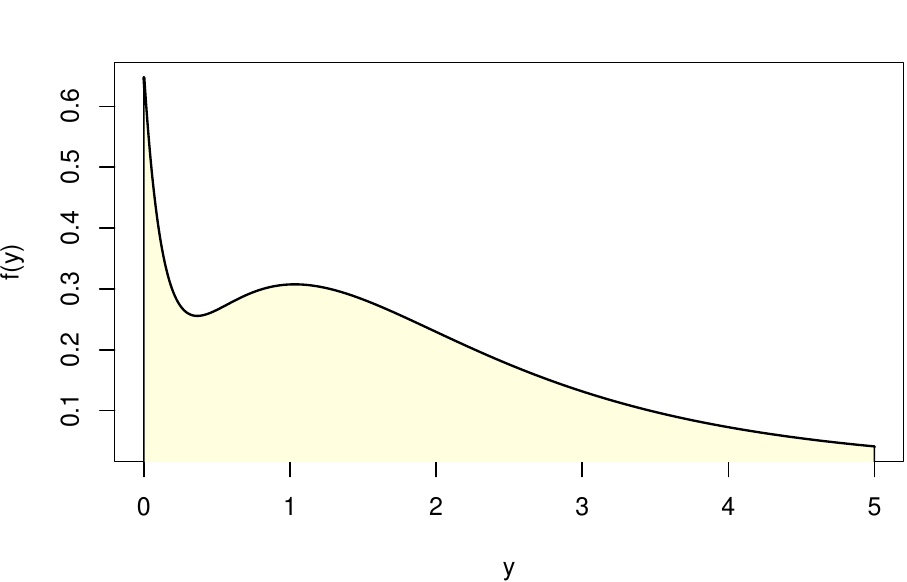}    \\
		\end{tabular}
		
	\end{center}	
	\caption{Four different examples of EGPD pdf (right panels) with $\xi=0.2$, and $\kappa=1,2,1,1$ (from  top to  bottom), in the left panels,  their corresponding 
		$b(u)$, see Equation \eqref{eq:mgpd_ML}.}
	\label{fig: example.pdf}
\end{figure}

A real-valued random variable is said to follow a generalized Pareto distribution (GPD) with an unit scale parameter if 
its cumulative distribution function (cdf) is defined by
$$
H_\xi(x)=\left\{
\begin{array}{lc}
	1-(1+\xi x)_+^{-1/\xi},&  \mbox{for } \xi\ne 0, \\
	1-\exp(-x), &\mbox{for } \xi = 0,
\end{array}
\right.
$$
where $\xi$ is the shape parameter  and $a_+ = \max(a, 0)$.
Its probability density function (pdf) is given by
$$
h_\xi(x)=\left\{
\begin{array}{lc}
	(1+\xi x)_+^{-1/\xi-1},&  \mbox{for } \xi\ne 0, \\
	\exp(-x), &\mbox{for } \xi= 0, 
\end{array}
\right.
$$
From the GPD quantile function, $H^{-1}_\xi(u)$, we can easily construct random draws of a EGPD. More precisely, let  $U$ be a uniformly distributed random variable on $[0,1]$. 
The  random variable $X$ defined as 
\begin{equation}\label{eq:X_def}
	X =  H_{\xi}^{-1}\left[ \left\{B^{-1}\left( U\right)\right\}^{1/\kappa} \right],
\end{equation}
where $\kappa$ and $\xi$ are  positive constants and $B(\cdot)$ represents any absolutely continuous  cdf on $[0,1]$, 
is said to follow a EGPD, denoted by $X \sim \mbox{EGPD}(\kappa, \xi, B)$,  if 
the pdf $b(u)$ of the random variable $B^{-1}(U)$ has positive  and finite end points, i.e. 
\begin{equation}\label{eq:constraints_on_b}
	0< b(0)  < \infty \mbox{ and } 0< b(1) < \infty.
\end{equation}
This definition appears slightly different from the one provided by \citet[Equation (5) in][]{Naveau2016}. But this is mainly a notation change 
as the cdf $$G(u)=B(u^\kappa)$$ was used   instead of $B(\cdot)$.  
In addition, a scale parameter called $\sigma$ was present in \cite{Naveau2016} but it is now included in the function $B$. Depending on the application at hand, $\sigma$ can be introduced but  parameter identifiability needs to be handled with care then. 
The notation with $G(\cdot)$ was a little bit ambiguous as the parameter $\kappa$ driving the lower tail was ``hidden'' in the cdf $G$ itself and this was not the case for $\xi$. 
In contrast,  
the new definition $\mbox{EGPD}(\kappa, \xi, B)$ makes a precise distinction among the role of $\kappa$ to model the lower tail,  the role of $B(\cdot)$ as the transfer function from low to heavy intensities and the role of $\xi$ for the upper tail.
From Equation \eqref{eq:X_def}, 
the cdf   of a $\mbox{EGPD}(\kappa, \xi, B)$ can be expressed, for all $x\geq 0$,  as 
\begin{equation}\label{eq: F bar}
	F(x)= B\left\{H_\xi(x)^\kappa\right\}.
\end{equation}
Concerning quantiles, we simply have, for all $u \in [0,1]$, 
\begin{equation*}
	F^{-1}(u)= H^{-1}_{\xi}\left\{  (B^{-1}(u))^{1/\kappa}\right\}.
	\label{eq:EGPD_pdf_cdf}
\end{equation*}
By taking the derivative of \eqref{eq: F bar} with respect to $x$, the pdf of a $\mbox{EGPD}(\kappa, \xi, B)$ can be written, for all $x \geq  0$,  as 
\begin{equation}\label{eq:mgpd_ML}
	f(x)= \kappa \cdot h_\xi(x) 
	\cdot H^{\kappa-1}_\xi(x) \cdot b\left\{ H^{\kappa}_\xi(x)\right\}.
\end{equation}
The right panels  of Figure \ref{fig: example.pdf} display four examples of $f(x)$ with  $\xi=0.2$. 
Each left panel shows the corresponding $b(u)$ used to generate the pdf $f(x)$. This figure illustrates the flexibility of the EGPD class.

Regarding the upper tail behavior of the survival function $\overline{F}(x)=P(X>x)$, the condition $0<b(1)<\infty$ in \eqref{eq:constraints_on_b} implies that
\begin{equation}\label{eq: b to b tilde}
	\lim_{x \to \infty} \frac{\overline{F}(x)}{\kappa \overline{H}_{\xi}(x)}=
	\lim_{x \to \infty} \frac{f(x)}{\kappa h_{\xi}(x)}= b(1).
\end{equation}
Hence, the upper tail behavior of a $\mbox{EGPD}(\kappa, \xi, B)$ is  equivalent to the one of a GPD with shape parameter $\xi$.
For small $x$,  the condition $0<b(0)<\infty$ in \eqref{eq:constraints_on_b}  gives 
us
\begin{equation}\label{eq:b0}
	\lim_{x \to 0} \frac{F(x)}{x^\kappa}=\lim_{x \to 0}\frac{f(x)}{\kappa x^{\kappa-1}}=    b(0).
\end{equation}
Thus, the parameter $\kappa$ controls the lower tail of $X$. 
So, the lower and upper tails comply with EVT on both sides. The parameters $\kappa$ and $\xi$  describe low and high values behaviors, respectively. 

The inference of the univariate EGPD parameters ($\kappa,\xi$) and the function $B$ will be treated in Section \ref{sec:inference}.
To close this section on univariate EGP distributions, we state a univariate result that will be useful for the extension into a multivariate context. 
\begin{lem}\label{lem:1/X}
	Let  $X \sim \mbox{EGPD}(\kappa, \xi, B)$ with $\kappa>0$ and $\xi>0$, then its inverse 
	$1/X$ is also a  EGPD with  $1/X \sim \mbox{EGPD}(1/\xi, 1/\kappa, \tilde{B})$  where  the pdf of $\tilde{B}$
	satisfies 
	$$
	\tilde{b}(0)=\kappa \; \xi^{1/\xi} \; b(1)\mbox{ and }\tilde{b}(1)=\xi \; b(0).
	$$
\end{lem}
The reader is invited to consult the appendix for the proofs of all lemmas and propositions.

\subsection{Multivariate extensions}

In the sequel, a vector of dimension $d$ is denoted  $\bx = (x_1,\ldots,x_d)\T$.
The bold symbols $\bzero$, $\bone$, and $\binfty$ refer to vectors all elements of which are equal to 0, 1, and $\infty$, respectively.
Operations between vectors such as addition and multiplication are to be understood componentwise. For instance, if $\bx = (x_1,\ldots,x_d)\T$  and $\ba= (a_1,\ldots,a_d)\T$, then $\bx - \ba$ is the vector with components $x_j - a_j$.  
Similarly, $\boldsymbol{1} /{\bx}$  is the vector with components $1/x_j$.

\begin{figure}[h!]
	\begin{center}
		\begin{tabular}{cc}
			\includegraphics[width=0.45\linewidth]{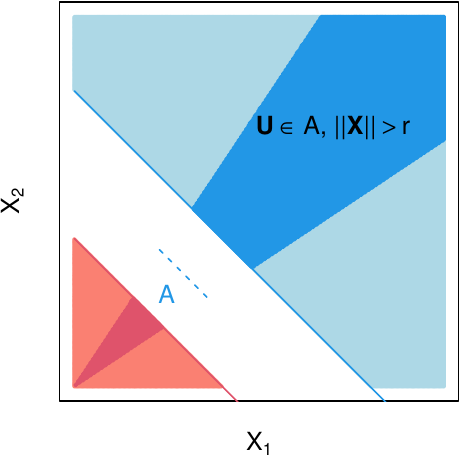}&
			\includegraphics[width=0.45\linewidth]{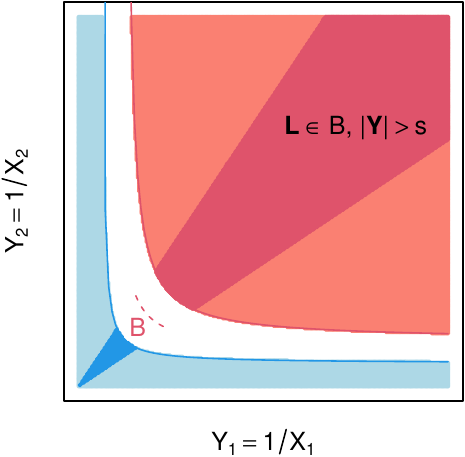}
		\end{tabular}
	\end{center}
	\caption{\textit{Left panel}. The blue set  represents an example of extremal region of interest for large values 
		as the sum $X_1+X_2$ is large, i.e.\ the radius $\Rad>r$.
		The dark blue set corresponds to the extremal region defined as  $\{\bU \in A,\Rad >r\}$ where $\bU= {\bX}/{\Rad}$ is called the angular component.
		\textit{Right panel}. For defining an extremal region for small values (red areas), we look at the set $\{X_1+X_2 < r\}$  whenever $r$ is small near zero. 
		By inverting the coordinates from $(X_1,X_2)$ to $\bY= (1/X_1,1/X_2)$ the set $\{X_1+X_2 < r\}$ of small values becomes 
		the set $\{|\bY| >s\}$ for the radius $|\bY|=1/(1/Y_1+1/Y_2)$.
		The dark red region defined by $\{\bL \in B,|\bY|>s\}$, with $\bL=\bY/|\bY|$, indicates the extremal region for the small values. 
	}\label{fig:extremes-2}
\end{figure}

For a positive random vector $\bX$ with identical heavy-tailed  margins,  MEVT 
indicates that a convenient way to represent upper tail behaviors in a $d$-dimensional space is to change the coordinate system with respect to  a given norm 
\citep[see, e.g.][]{dehaan:ferreira:2006,beirlant:goegebeur:teugels:segers:2004}.
With this norm,   the random vector of interest can be  decomposed into two parts: 
its  ``radial"  component $\Rad$ and  the ``angular" component $\bU=(U_1,\ldots,U_d)$,  with  $U_i ={X_i}/{\Rad}$,  $i=1,\ldots,d$, namely
\begin{equation}\label{eq:modelRW}
	\bX = \Rad \times \bU, 
\end{equation}
where $\left\|\bU \right\| =1$. This representation has a few advantages. 
Most importantly, it is easy to define large extreme events as occurrences of large radii.
For example, for the bivariate case with 
$\norm{\bx}=x_1 +  x_2$, the blueish  region in the left panel of Figure \ref{fig:extremes-2} displays bivariate extremes whenever the radius $\Rad$ is large. 
The complex task of defining multivariate extremes becomes, albeit the choice of the norm,  a classical univariate problem.  
The EGPD distribution defined by \eqref{eq:X_def} will be the  building block to model $\Rad$.
The radial and angular representation is closely linked to the concept of   multivariate regular variation 
that has  been extensively studied   by the extreme value community \citep[see, e.g.][for a recent review]{NaveauSegers25}. In nutshell, a multivariate regular variation distribution obeys the following  conditions 
\begin{equation}\label{ref: mevt conditions}
	\begin{cases}
		& \mbox{$\Rad$ independent of $\bU$ when $\Rad$ gets  large},\\
		& \Pr( \bU \in A \; | \;  \Rad>x) \mbox{ has a non-degenerate limit as } x \to \infty.
	\end{cases}
\end{equation}
The first condition tells us that the strength of an extreme event is given by the heavy-tailed distributed $\Rad$ those large intensity values do not depend on $\bU$. 
Given that such intensities are large, the second condition indicates that the dependence structure is entirely captured by the angular component. 
These conditions are closely linked to the so-called concept of max-stable domain of attraction used to describe multivariate block maxima distributions \citep[see, e.g.][]{dehaan:ferreira:2006}. 
For example, all extreme value copula (after adequately transforming their marginals) belong to this domain and satisfies conditions \eqref{ref: mevt conditions} \citep[see, e.g.][]{beirlant:goegebeur:teugels:segers:2004}. 
Concerning low extremes, the focus moves to the component-wise inverse vector of  $\bX$, say $\bY=\mathbf{1} / \bX$ and the multivariate regular variation has to applied with respect to the  transformed risk functional defined by
$$
\homog{\bz}=\frac{1}{\norm{\boldsymbol{1} /\bz}},\quad \mbox{for } \bz \mbox{ such that } \min z_i>0,
$$
as the set of large $\bY$, i.e. 
$$
\{|\bY| > y \} = \left\{\frac{1}{\norm{\boldsymbol{1}/\bY}} > y \right\}= \left\{x > \norm{\bX} \right\}
$$
is equal to the set of small $\bX$ (low extremes at the original scale).
Although $\homog{\bx}$ is not a norm, it is, as $\norm{\bx}$, homogeneous of order one, i.e. 
$
\homog{s{\bx}} = s \homog{\bx} \mbox{ for any positive scalar $s$}. 
$
For example, we can visualize  the bivariate case with $\norm{\bx}=x_1+x_2$ in Figure  \ref{fig:extremes-2}.
Moving from  the left panel to  the right panel  allows the reader to see how the regions of interest (low and high extremes)  are flipped when 
the   component-wise inverse vector $\bY$ are chosen instead of $\bX$ (right panel).
More precisely, the left panel provides a two-dimensional  regions to which a  MEVT model assigns probability.
i.e.  the blue set defined by $\{\bU \in A,\Rad >r\}$. 
Then, the set $\{X_1+X_2 < r\}$  for small $r$ becomes 
the set $\{1/X_1 +1/X_2>s\}$ for large  $s=1/r$, see the right  panel.

A key aspect of this work is to define a multivariate random vector with EGPD stability properties and MEVT compliancies for low and large extrmes, i.e. for both $\norm{\bX}$ and $|\bY|$. 
In terms of stability, one desiderata is that the marginals of a EGPD vector should be EGPD.
In addition, it would handy if the radius $\Rad$ in the polar decomposition remains EGPD. 
Concerning low extremes,    EGPD stability  will be welcome for the appropriate functional, i.e. expressed in terms of $\bY=\mathbf{1} / \bX$ and $|\bY|$.   
The following definition lists the needed requirements to define   a multivariate EGPD vector with such properties. 

\begin{definition}\label{def: MEGPD}
	Let $\bX$ be a positive multivariate random vector of dimension $d$.
	We say that $\bX$ follows a multivariate EGPD with respect to the norm $\norm{.}$ if   the  following conditions are satisfied.
	\begin{enumerate}
		\item Each marginal   is identically EGPD distributed  with   $X_i \sim \mbox{EGPD}(\kappa, \xi, B)$  for some  $\xi>0$ and $\kappa>0$.
		\item 
		There exists a  positive and finite constant $c_+$ such that, for all $i=1,\ldots,d$, 
		\begin{equation}\label{eq: tail equivalence conditions upper}
			\lim_{x \to \infty } \frac{\Pr( \Rad > x) }{\Pr( X_i > x) }=c_+.
		\end{equation}
		\item 
		There exist two positive and finite constants, $a$ and   $c_-$  such that, for all $i=1,\ldots,d$,  
		\begin{equation}\label{eq: tail equivalence conditions lower}
			\lim_{x \to  0 } \frac{\Pr( \Rad \leq  x) }{\{\Pr( X_i \leq  x) \}^a}=c_-. 
		\end{equation}
	\end{enumerate}
\end{definition}
The first requirement that imposes    a EGPD for each marginal  is natural. 
The second and third conditions are necessary to insure that high and low extreme intensities captured by  $\norm{\bX}$ and $|\bY|$ respectively, are also EGPD, see Lemma 
\ref{lem: norm(X) EGPD} below. 
In terms of applicability, the question is to determine if conditions \eqref{eq: tail equivalence conditions upper} and \eqref{eq: tail equivalence conditions lower} are often satisfied in practice. 
In the bivariate case with $\Rad=\max(X_1,X_2)$, condition \eqref{eq: tail equivalence conditions upper} is equivalent to have $c_+=2-\chi$ where $\chi$ is the classical tail dependence coefficient, i.e. the limit of $\Pr( X_2 > x | X_1 >  x)$ for large $x$.
If $\bX=(X_1,X_2)\T$is a max-stable vector with $\Pr( \max(X_1,X_2)\leq x)=\Pr(X_1\leq x)^a$ for some $0<a<1$, then condition \eqref{eq: tail equivalence conditions lower} is satisfied for $\Rad=\max(X_1,X_2)$ with $c_-=1$.

In this work, we will focus on the sum-norm $\Rad=X_1+\ldots+X_2$.
In the special case when  all $X_i$ are independent, the constants can be obtained explicitly for this norm.
In particular, we get $c_+=d$ in condition \eqref{eq: tail equivalence conditions upper}  by the so-called Feller lemma \citep[VIII.8]{feller:1971} that deals with the sum of independent heavy-tailed distributed random variables. 
For dependent cases, extensions of Feller lemma have been proposed and, under general multivariate regular variation in the upper tails, \eqref{eq: tail equivalence conditions upper} is satisfied for most MEVT models; see \cite{fougeres:mercadier:2012} and Chapter 4 of 
\cite{KulikSoulier20} for details.
Concerning \eqref{eq: tail equivalence conditions lower}, one can notice that the pdf of $\Rad$ corresponds to a convolution of the same pdf. Classical convolution and Taylor approximation arguments lead to $a=d$ in the  independent  case. 
For   the case of regular variation, one can show that $a=1$ \citep{KulikSoulier20}. 
These examples explain the need for the nonfixed constant $a$ in \eqref{eq: tail equivalence conditions lower}.  

Knowing now that   the class of models that satisfy \eqref{eq: tail equivalence conditions upper}    and \eqref{eq: tail equivalence conditions lower} is large, our next task is to explore what are the main properties of models that satisfy Definition \ref{def: MEGPD}.

\begin{lem}\label{lem: norm(X) EGPD}
	Let $\bX$ satisfying Definition \ref{def: MEGPD}. 
	Then, there exists a  cdf $B_d$ such that the radius 
	$\Rad \sim \mbox{EGPD}(a \kappa, \xi, B_d)$
	with 
	$$
	b_d(0)=  c_- \cdot b(0)^a  \mbox{ and } b_d(1)= c_+ \cdot b(1)/a.
	$$
	Consequently, $\homog{\bY}={1}/{\norm{\boldsymbol{1}/\bY}}={1}/{\norm{\bX}}$ also follows a 
	$\mbox{EGPD}(1/\xi, 1/(a\kappa), \tilde{B}_d)$ with 
	$$
	\tilde{b}_d(0)=\kappa \; \xi^{1/\xi} \; b_d(1)\mbox{ and }\tilde{b}_d(1)=\xi \; b_d(0).
	$$
\end{lem}
This lemma indicates that the upper  tail indicators, $ \xi$, remains unchanged 
from marginal behaviors to $\Rad$, while  the lower tail behavior can be have a changing $\kappa$.

A key aspect in Definition \ref{def: MEGPD}  is that the radius and the angular component can be dependent in the bulk of the distribution of $\bX$. This allows a large flexibility in terms of modeling.  
Another point is that the study of the small values of $\bX$ can be also handled by MEVT if we study the large values of $\mathbf{1} / \bX$, via 
$$
\bY =    |\bY|  \times \bL, 
$$
where $\left|\bL \right| =1$.

To summarize, the three main differences with classical MEVT modeling are that 
\begin{itemize}
	\item[i)] our interest is not only on the upper extremal behavior of $\bX$, but also in the bulk and the  lower extremal behavior; 
	\item[ii)]  the radial component  follows  a EGPD, and consequently be in compliance with EVT for both small and large values of $\Rad$ and $|\bY|$;
	\item[iii)] for moderate values, the radial component is not necessarily assumed independent of the angular component for the bulk. In particular, the degree of dependence will change according to the value $\Rad$. 
\end{itemize}

\subsection{Multivariate EGPD logistic-heteroscedastic modeling}\label{subsec:biv_mod}
To further illustrate our modeling approach, we  focus on a specific  case. For a discussion of more general cases, please refer to the concluding section.

As we do not want to assume $\bU$ and $\Rad$ independent for moderate values of $\Rad$, a conditional model is needed to characterize this link. A trade-off is needed to balance modeling flexibility, inferential scarcity and computational limits. 
Inspired by the literature on logistic regression and compositional analysis  we focus here on the simple transformation where other coordinates  are compared with respect to an arbitrarily chosen  component (here the $d$ component) via the log-ratio $\log\left({X_i}/{X_d}\right)$.

Note that, by construction, this ratio is not affected by $||\bX||$ and 
$
\log\left({X_i}/{X_d}\right)= \log\left({U_i}/{U_d}\right)
$.
For example, in our trivariate case in hydrology (see Application Section), we  will   model three runoff time series  on the same river  by assuming that    the two upstream location   impact the two downstream locations. So, the upstream location will correspond to the d-coordinate and ratios of the type ${U_i}/{U_d}$ will be modeled. 
This  log-ratio  representation is well known in compositional analysis. 
Still, for low and large extremes, the effect of $\Rad$ should disappear on $\bU$ in MEVT regular variation based  setups. 
To deal with issue, we impose a heteroscedastic factor that will disappear with low and large extremes. More precisely, we define the following MEGPD vector.

\begin{definition}\label{def: BEGPD}
	Let $\bX = \Rad \times \bU$  be a   MEGPD random vector that satisfies Definition \ref{def: MEGPD}.
	We say that $\bX$ follows a   logistic-heteroscedastic EGPD   if   
	the following  log-ratio, given the radius $\Rad=r$,
	can be expressed  as 
	\begin{equation}\label{eq:d-dim_model}
		\log\frac{X_i}{X_d}=  \delta(r)\, Z_i,  \;\mbox{for $i=1,\ldots, (d-1)$,}
	\end{equation}
	where  the (d-1) dimensional vector  $\bZ=(Z_1,\ldots, Z_{d-1})\T$ is a zero-mean exchangeable  random vector independent of $\Rad$ and  
	$\delta(\cdot)$ is a positive measurable function  such that,  
	uniformly on any compact of the real line,  
	\begin{equation}\label{eq: unif conv}
		\lim_{r \to 0}   \delta(r)  = \delta(0)\qquad    \mbox{and}\qquad
		\lim_{r \to \infty}  \delta(r) =  \delta(\infty), 
	\end{equation}
	for some finite positive constants $\delta(0)$ and $\delta(\infty)$.
\end{definition}
From Lemma \ref{lem: norm(X) EGPD}, we know that the radius follows an EGPD.
Clearly, Equation \eqref{eq:d-dim_model} contains a logistic transformation. 
The scaling term $\delta(\cdot)$ can be viewed as heteroscedasticity, which drives the dependence structure. 
Given $\Rad=r$,  we can write
$$
U_i = \cfrac{e^{\delta(r)Z_i}}{1 + \sum_{j=1}^{d-1} e^{\delta(r)Z_j}}, \quad \text{for } i = 1, \ldots, d-1,\qquad U_d = \cfrac{1}{1 + \sum_{j=1}^{d-1} e^{\delta(r)Z_j}},
$$

and the distributional symmetry of $Z$ implies that, given $\Rad=r$, we have $\E( U_i |\Rad=r) = 1/d$ for all $r$.  
Concerning the lower tail of $\bX$, i.e. $\bY=\mathbf{1}/\bX$, one can note that 
$$
\log \left(\cfrac{Y_i/|\bY|}{Y_d/|\bY|}\right)  = 
\log\frac{X_d}{X_i} = - \log\frac{X_i}{X_d}.
$$
Conditionally on small $\Rad$ (i.e. on large $|\bY|$), the symmetry of $\bZ$ in definition \eqref{eq:d-dim_model}  implies that 
the lower tail dependence, via $- \log(X_i/X_d)$, is driven by the values of $\delta(\Rad)$ when $\Rad$ is small, i.e. $\delta(0)$.

In terms of interpretation, it is clear from \eqref{eq:d-dim_model} that if $\delta(\Rad)$ remains constant for large values of $\Rad$, then 
we are in the regular variation setup in which large values of  $\Rad$ does not impact the angular component (see Proposition \ref{prop:convergence}).

Concerning the function $\delta(\cdot)$ and its left and right limits, condition
\eqref{eq: unif conv} needs to be refined. 

\begin{prop}\label{prop:convergence}
	The pdf of the MEGPD model from Definition \ref{def: BEGPD} can be written as 
	\begin{equation}\label{eq:pdf_logistic_model}
		f(\bx) = \frac{1}{\delta^{d-1}(\norm{\bx} )} 
		f_{\bZ}\left(\frac{1}{\delta(\|\bx\|)}(\log(x_1/x_d),\ldots, \log(x_{d-1}/x_d))\right\} 
		f_{\norm{\bX}}(\norm{\bx} ) 
		\frac{\norm{\bx} }{\prod_{i=1}^{d} x_i},
	\end{equation}
	where $f_{\bZ}(\cdot)$ is the pdf of $\bZ$ and $f_{\norm{\bX}}(\cdot)$
	is the pdf of $\mbox{EGPD}(\kappa, \xi, B)$.
	
	\noindent If   the following uniform continuity holds 
	\begin{equation}\label{eq: pdf + }
		\lim_{t \rightarrow \infty} \sup_{\norm{\bx}=1}   \Bigg|  
		\frac{f(t \bx)}{t^{-d-1/\xi}} -\lambda_+(\bx)  \Bigg|=0
	\end{equation}
	with 
	
	$$
	\lambda_+(\bx)=\frac{b(1)\xi^{-1/\xi-1}}{\delta^{d-1}(\infty)}   f_{\bZ}\left(\frac{\log(x_1/x_d), \ldots, \log(x_{d-1}/x_d)}{\delta(\infty)}\right)  \frac{\norm{\bx}^{-1/\xi}}{\prod_{i=1}^{d}x_i},
	$$
	for any $\bx \in [\bzero,\infty) \setminus \{\bzero\}$, then the upper tail of $\bX$ is regularly varying with 
	$$
	1-F(t \bx) \sim t^{-1/\xi} \int_{[\bzero,\bx]^c} \lambda_+(y) dy.
	$$
	
\end{prop}

The proof the  proposition is based on regularly functions results for densities, 
\cite[e.g., see Theorem 6.4 in][]{Resnick:2007}.
The upper tail  equivalence expressed with $1-F(t \bx)$ 
indicates that, as expected in classical MEVT, the component driving the upper tail dependence becomes independent of the radial component as the radius becomes large.
The same type of results can be derived  for the lower upper tail case when the radius becomes close to zero. 

One convenient choice for the  distribution  $f_{\bZ}(\cdot)$ in \eqref{eq:pdf_logistic_model} is the standardized 
multivariate  Gaussian distribution with constant pairwise correlation $-1<\rho<1$.
With this choice,  the density function in  \eqref{eq:pdf_logistic_model} can be written as 
\begin{eqnarray}\label{eq:gaussian_pdf_logistic_model}
	f(\bx)&=&\kappa\,  h_\xi(\|\bx\|) 
	H^{\kappa-1}_\xi(\|\bx\|) b\left( H^{\kappa}_\xi(\|\bx\|)\right) \times \nonumber
	\\
	&&  (2\pi)^{-(d-1)/2}\delta(\|x\|)^{-(d-1)}\mathrm{det}(\bC)^{-1/2}\exp\left\{-\cfrac{1}{2}\,\cfrac{\bv\T\bC^{-1} \bv}{\delta(\|\bx\|)^2} \right\} \cfrac{\|\bx\|}{\prod_{i=1}^dx_i},
\end{eqnarray}
where $\bv=(\log(x_1/x_d),\ldots,\log(x_{d-1}/x_d))\T$ and $\bC=(c_{ij})$ is a $(d-1) \times (d-1)$ correlation matrix, with $c_{ij}=\rho$ for $i\ne j$.

\begin{figure}
	\begin{center}
		\begin{tabular}{ccc}
			\multicolumn{3}{c}{\includegraphics[width=0.6\linewidth]{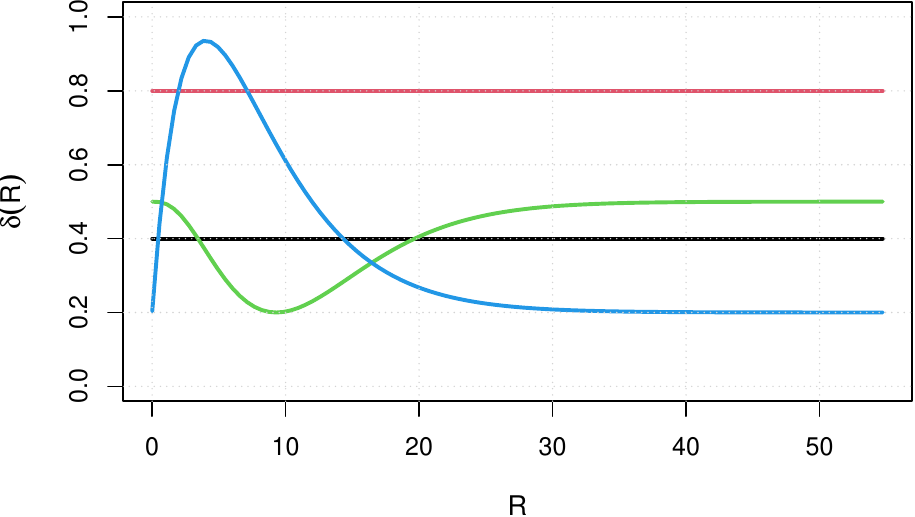}}\\
			\includegraphics[width=0.33\linewidth]{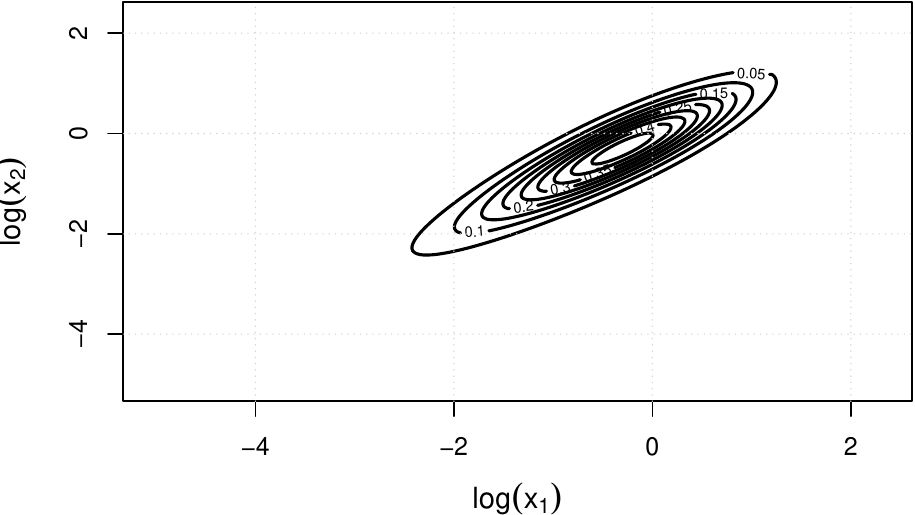} &
			\includegraphics[width=0.33\linewidth]{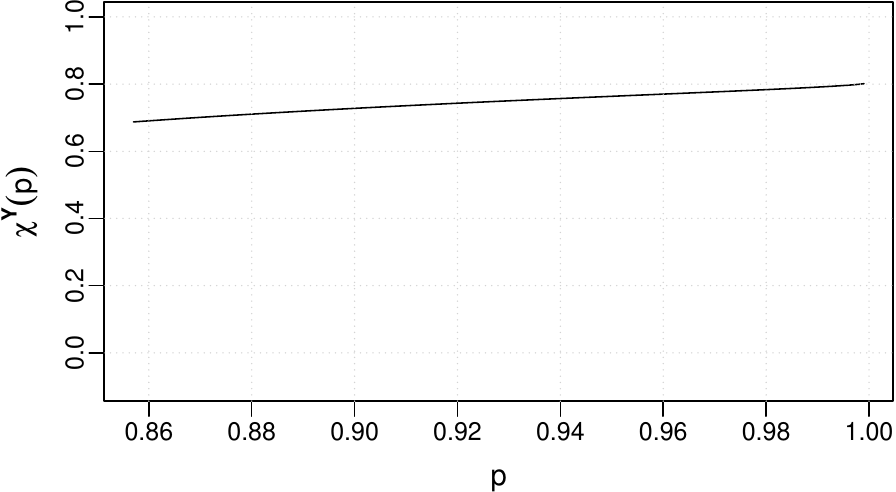} &
			\includegraphics[width=0.33\linewidth]{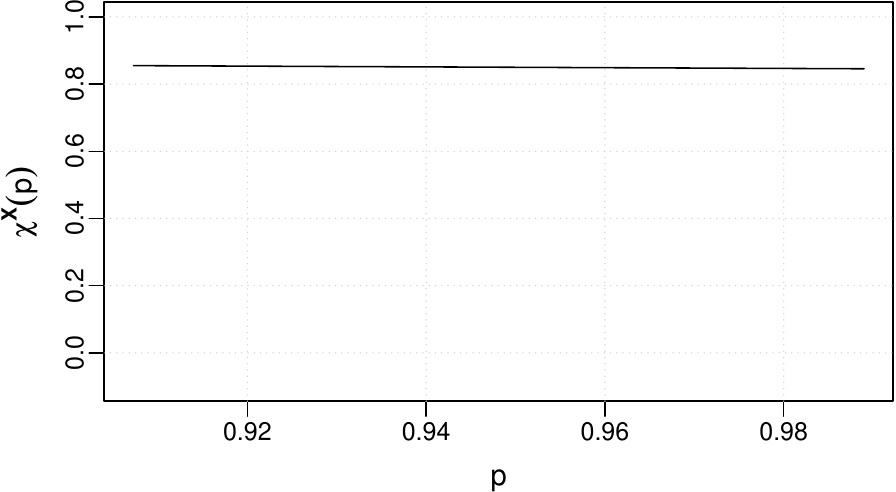} \\
			\includegraphics[width=0.33\linewidth]{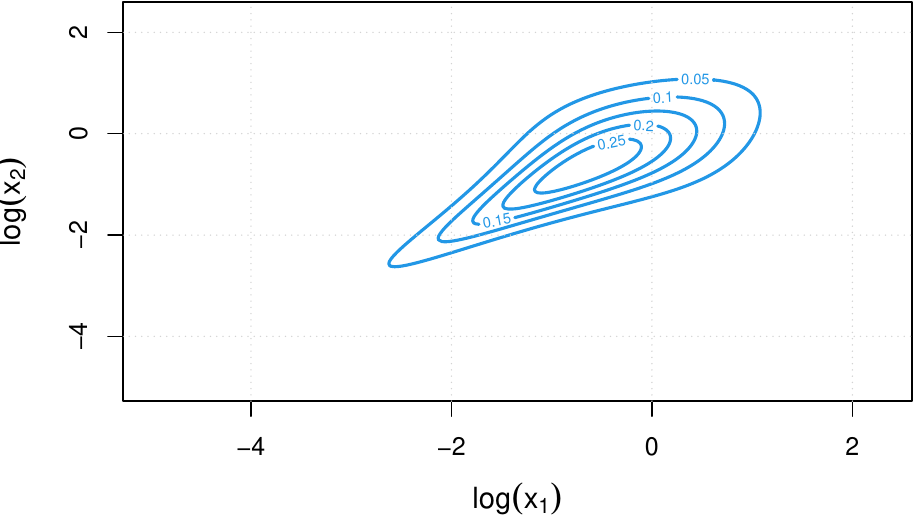} &
			\includegraphics[width=0.33\linewidth]{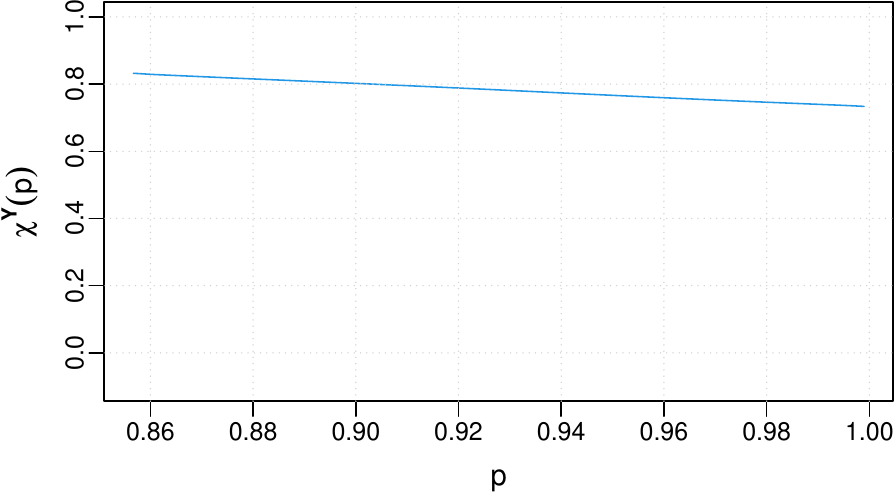} &
			\includegraphics[width=0.33\linewidth]{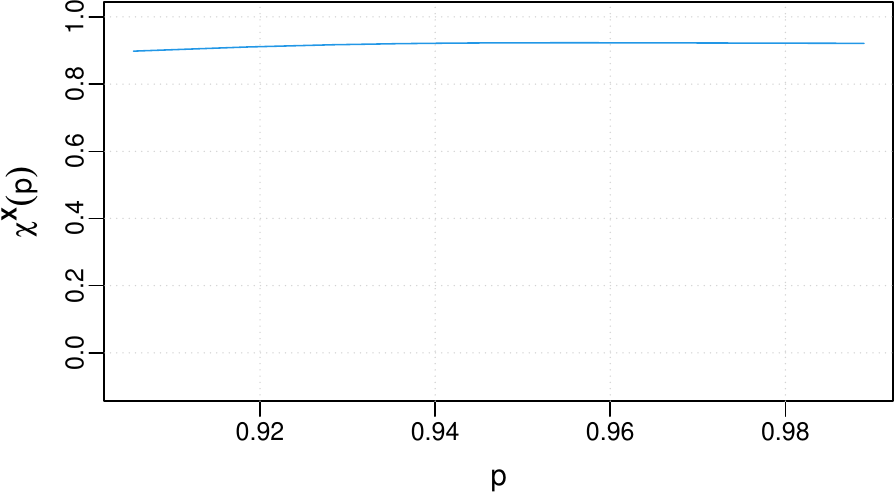} \\
			\includegraphics[width=0.33\linewidth]{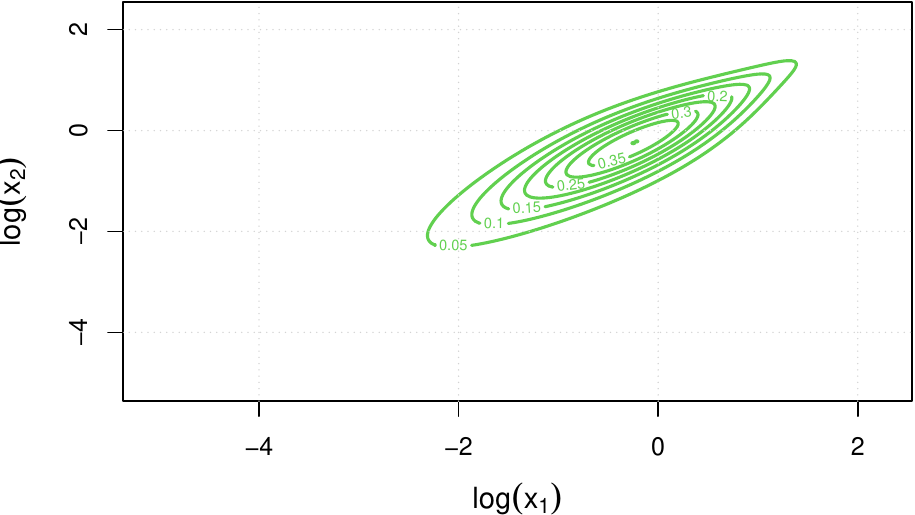} &
			\includegraphics[width=0.33\linewidth]{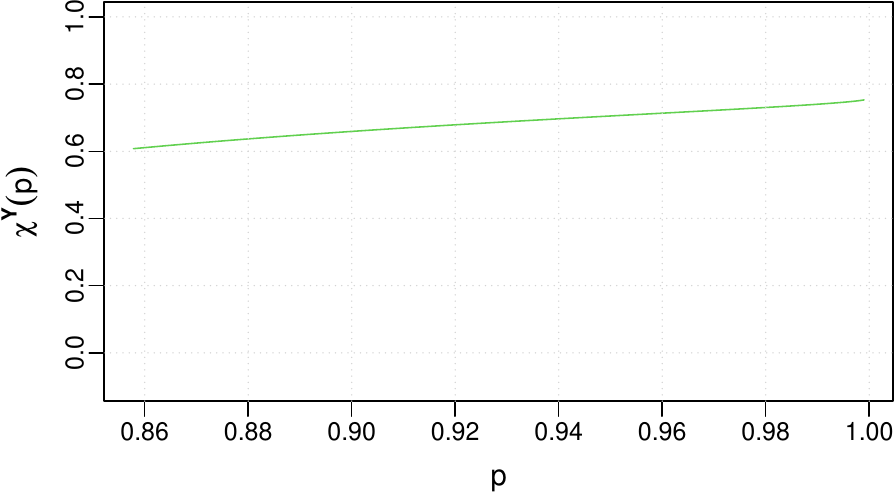} &
			\includegraphics[width=0.33\linewidth]{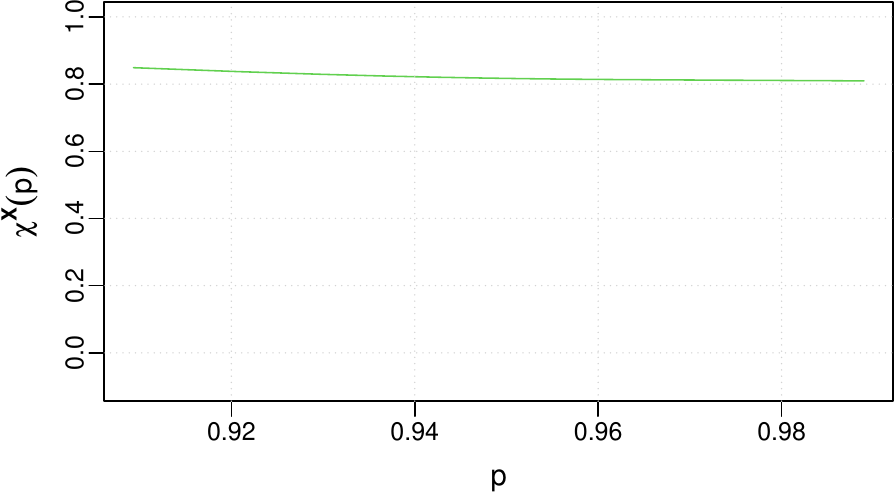} \\
			\includegraphics[width=0.33\linewidth]{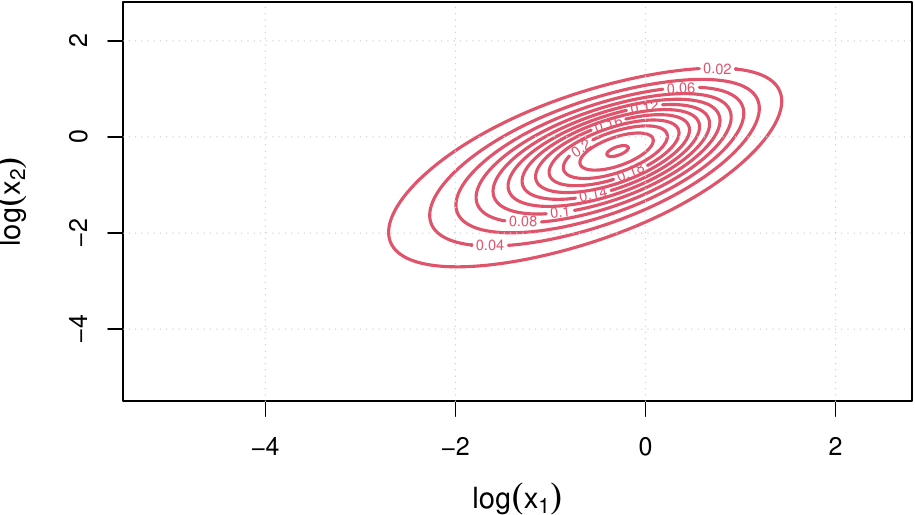} &
			\includegraphics[width=0.33\linewidth]{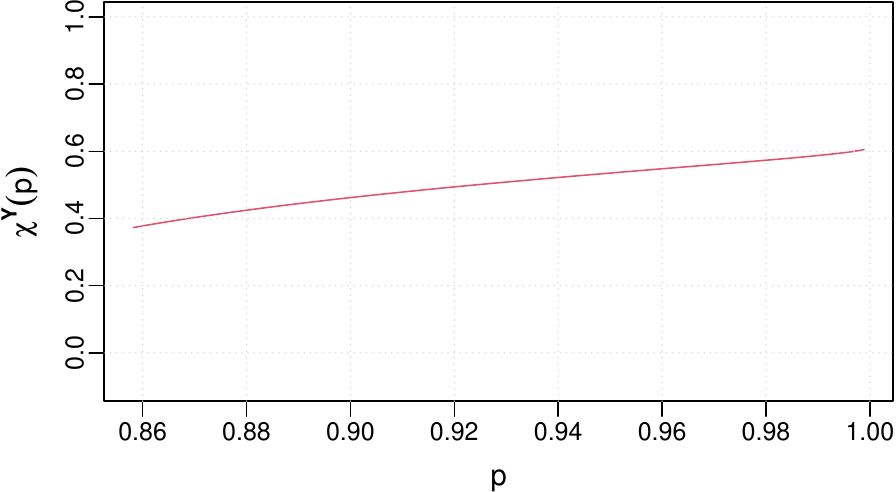} &
			\includegraphics[width=0.33\linewidth]{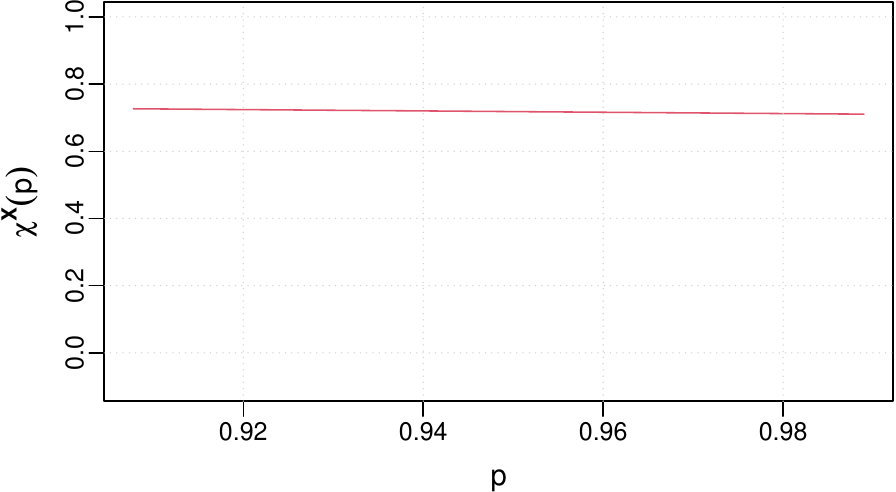} \\
		\end{tabular}
	\end{center}
	\caption{Top: Four examples of the standard deviation function $\delta(\cdot)$ (Black, Blue, Green and Red color). Left column: contour plots of the corresponding  bivariate density as in \eqref{eq:biv}; center and right columns: plot of the extremal dependence functions $\chi^{(\bY)}(p)$ and $\chi^{(\bX)}(p)$ as defined in \eqref{eq:chiu}}
	\label{fig:bivariate-distribution}
\end{figure}

In this setting, we will look at how useful and flexible the function $\delta(\cdot)$ is for modeling the dependence  between two variables. The density function for the pair of random variables $X_1$ and $X_2$ of model \eqref{eq:pdf_logistic_model} can be simplified to
\begin{eqnarray}\label{eq:biv}
	f(x_1,x_2)&=&\kappa  h_\xi(x_1+x_2) 
	\cdot H^{\kappa-1}_\xi(x_1+x_2) b\left( H^{\kappa}_\xi(x_1+x_2)\right) \times \nonumber
	\\
	&&  \cfrac{1}{\sqrt{2\pi}\delta(x_1+x_2)}\exp\left\{-\frac{1}{2}\left[\cfrac{\log(x_1/x_2)}{\delta(x_1+x_2)}\right]^2\right\} \frac{(x_1+x_2)}{x_1x_2},
\end{eqnarray}
Figure \ref{fig:bivariate-distribution} illustrates the flexibility
through four examples. Each colored curve in the top panel represents a different function  $\delta(\cdot)$. The first three-row column displays  the four associated bivariate densities. 
We set $\xi=0.1$, $\kappa=2$ and $b(u)= I_{(0,1)}$, i.e.  $B(u)=u I_{(0,1)}$   that corresponds to the model 1 in \citet{Naveau2016}.
Additionally, each   middle   (right) panel shows  the  lower (upper) extremal dependence coefficients defined as 
\begin{equation}\label{eq:chiu}
	\chi^{(\bY)}(p)=\cfrac{\P(F_{Y_1}(Y_1)> p,F_{Y_2}(Y_2)> p)}{1-p}, \quad \chi^{(\bX)}(p)=\cfrac{\P(F_{X_1}(X_1)> p,F_{X_2}(X_2)> p)}{1-p},
\end{equation}
for $0<p<1$, with  $\bY=(Y_1,Y_2)=(1/X_1,1/X_2)=\boldsymbol{1}/\bX$.

\section{Inference}\label{sec:inference}

Suppose that we observe $n$ i.i.d realizations $\bx_i$, $i=1,\ldots,n$ of $\bX$ with density \eqref{eq:gaussian_pdf_logistic_model}.
The inference for $\kappa,\xi, \,b(\cdot)$ and $\delta(\cdot)$ can easily be done in two steps by considering the transformed observations $\|\bx_i\|,\,\log(x_{i,1}/x_{i,d}),\ldots,\log(x_{i,d-1}/x_{i,d})$ and exploiting the hierarchical structure of the model \eqref{eq:d-dim_model}.

\paragraph{First step: radial component EGPD}
In the first step we estimate the parameter of the distribution \eqref{eq: F bar} of $\Rad$. We estimate the density $b(\cdot)$ using a Bernstein basis polynomial of degree $m+1$, as described in \citet{Vitale:1975}, i.e.
\begin{equation}\label{eq:Bernstein}
	\widehat{b}_m(u)= \sum_{k=1}^m\omega_{k,m}\beta_{k,m-k+1}(u)
\end{equation}
where $\beta_{i,j}(u) \mathrel {=} {\binom {j}{i }}u^{i }\left(1-u\right)^{j-i}$. Here the weights $\omega_{k,m}$ are equal to $\omega_{k,m}=\mathbb{B}_n(k/m)-\mathbb{B}_n((k-1)/m)$ where $\mathbb{B}_n$ denotes the empirical cumulative distribution function of a random sample of size $n$ with distribution $B$.
Such a sample $u_i$, $i=1,\ldots,n$ can be derived by  by assuming knowledge of the values  of the parameters  $\kappa$ and $\xi$ and setting $u_i=H_\xi(\|\bx_i\|)^\kappa$. 

By plugging $\widehat{b}_m$ into \eqref{eq:mgpd_ML}, an estimate of $\kappa$ and $\xi$ can be obtained by maximising the log-likelihood
$$
l(\kappa,\xi)= \sum_{i=1}^n \left\{\log\kappa+ (\kappa-1)\log H_\xi\left(\|\bx_i\|\right)+\log h_\xi\left(\|\bx_i\|\right)+\log \widehat{b}_m(H_\xi(\|\bx_i\|)^\kappa)\right\}.
$$
A similar estimation procedure for estimating $\kappa$ and $\xi$ has been used by \citet{Tencaliec:Favre:Naveau:Prieur:Nicolet:2020}, by means of probability-weighted moments. 
Two critical points appear in this step.

\begin{enumerate}
	\item 
	The choice of the basis size $m$ that corresponds to resolving a bias-variance trade-off. \citet{babu:canty:chaubey:2002} showed that $m$ should be of the order $n/\log(n)$ for consistent convergence results and  our numerical experiences indicate that $m=\lfloor 0.5\, n/\log(n)\rfloor$ works well. Alternatively cross-validation techniques, as in \citet{Tencaliec:Favre:Naveau:Prieur:Nicolet:2020}, can help but increase the computational burden. 
	\item Concerning the constraints  $b(0) >0$ and $b(1) >0$,  these are satisfied when  $\omega_{1,m}=\mathbb{B}_n(1/m)$ and  $\omega_{m,m}=1-\mathbb{B}_n(1-1/m)$ are positive.
	A simple but effective strategy we have implemented to guarantee the bounds in the estimation is the following. 
	Let $k^*$ be the first integer such that $\omega_{k,m}>1/m$.
	If $k^*=1$, keep $\omega_{1,m}$ but otherwise, set $\omega_{1,m}=1/m$ and $\omega_{k^*,m}= \omega_{k,m}-1/m$. 
	For the case $\omega_{m,m}=0$, taking the last positive value, a similar idea applies.

\end{enumerate}

\paragraph{Second step.} 
To estimate $\delta(\cdot)$ in the conditional distribution 
\eqref{eq:d-dim_model} of $\log(X_i/X_d)$, $i=1,\ldots,d-1$,  given $\Rad=r$, we follow a penalized likelihood approach \citep{Wood2017} in which we represent  $\log \delta(r)$  as a  linear combination of $K$ basis functions, such as cubic splines, $S_j(\cdot)$, $j=1,\ldots,K$ namely 
\begin{equation}\label{eq:link}
	\log \delta(r) :=h(r;\bgamma)=\gamma_0+\sum_{j=1}^K \gamma_j S_j(r),\qquad \bgamma=(\gamma_0,\ldots,\gamma_K)\T    
\end{equation}
and we maximize the (penalized) log-likelihood, with respect $\rho$ and $\bgamma$, 
\begin{equation}\label{eq:penlik}
	pl(\rho,\bgamma,\lambda)=-\sum_{i=1}^n\left\{(d-1)
	h(\|\bx_i\|;\bgamma)+\log(\mathrm{det}(\bC))  +\cfrac{1}{2}\cfrac{\bv_i\T \bC^{-1} \bv_i}{\exp( 2h(\|\bx_i\|;\bgamma))}\right\}+\lambda \bgamma\T \bP \bgamma,
\end{equation}
where $\bv_i=(\,\log(x_{i,1}/x_{i,d}),\ldots,\log(x_{i,d-1}/x_{i,d}))$.

Here the smoothing parameter $\lambda$ is a fixed positive scalar and the  $\bP$ is a positive semi-definite matrix. The entries of the penalty matrix $\bP$ are the integrals of the products of the second derivatives of pairs of cubic spline functions, see \citet[][Section 5.3]{Wood2017} for more details.
In this framework the estimation of  $\bgamma$  and the selection of the amount of smoothing, i.e. $\lambda$, can be easily accomplished with standard software like the   R package \texttt{mgcv} following  Algorithm \ref{algo:1}.

\begin{algorithm}
		\caption{Iterative Maximization Algorithm for \eqref{eq:penlik}}\label{algo:1}
		\begin{algorithmic}[1]
			\REQUIRE  Initial guess $\rho^{(0)}$, convergence threshold $\epsilon$.
			\STATE $k \gets 0$
			\REPEAT
			\STATE Fix $\rho^{(k)}$, compute $(\bC^{(k)})^{-1/2}$ the Cholesky decomposition of $(\bC^{(k)})^{-1}$  and 
			$\bv_i^{(k)}=(\bC^{(k)})^{-1/2}\bv_i$, for $i=1,\ldots, n$
			\STATE Using $\bv_i^{(k)}$, get  the estimates 
			$\bgamma^{(k)}$ and $\lambda^{(k)}$ using a REML criterion as in \citet[][Section 6.2]{Wood2017}
			
			\STATE Compute profile function $f(\rho) = l_v(\rho, {\bgamma}^{(k)},{\lambda}^{(k)})$
			\STATE Update $\rho^{(k+1)}$ using optimization (e.g., Newton-Raphson or gradient ascent) on $f(\rho)$
			\STATE $k \gets k + 1$
			\UNTIL{$|\rho^{(k)} - \rho^{(k-1)}| < \epsilon$}
			\RETURN $\hat{\rho} = \rho^{(k)}$ and $\hat{\bgamma} = \bgamma^{(k)}$.
		\end{algorithmic}
	\end{algorithm}

\section{Numerical examples}\label{sec:numerical_examples}
\subsection{Bivariate copula examples} 
The objective is to illustrate whether the semi-parametric formulation specified in Section \ref{subsec:biv_mod} is sufficiently flexible to accommodate the dependence of three copula examples commonly used in EVT and previously examined in \citet{Wadsworth:Tawn:Davison:Elton:2017}.
According to Sklar’s Theorem \citep{Sklar:1959}, the bivariate distribution function  of a continuous random vector $\bX=(X_1,X_2)\T$ can be written as the composition of a copula, $C:[0,1]^2\rightarrow [0,1]$
as follows
$$
F(\bx)= C(F_1(x_1),F_2(x_2)),
$$
where   the margins $F_1$ and $F_2$  of $\bX$ are assumed to be  EGPD   with 
$\xi=0.1$, $\kappa=2$ and $b(u)= I_{(0,1)}$ and  three different dependence structures with copulas derived by
\begin{itemize}
	\item[(a)] the bivariate  symmetric logistic distribution \citep{coles:tawn:1991}  defined (on the Fr\'echet scale) as  
	$$\exp\left\{- \left(\frac{1}{y_1^{1/\alpha}}+\frac{1}{y_2^{1/\alpha}}\right)^\alpha \right\}, \quad 0<\alpha\leq 1; $$
	\item[(b)] the inverted bivariate  symmetric logistic distribution,  namely  the distribution of the component-wise inverse of random variables following  (a);
	\item[(c)] the standardized bivariate Gaussian distribution with correlation $1-\alpha$,  $0<\alpha\leq 1$.
\end{itemize}	
In our examples we set $\alpha=0.2$. Note that the first copula is asymptotically dependent in the upper tails i.e. the limit of $\chi^{(\bX)}(p)$ is positive for $p\rightarrow 1$, and the second and the third copulas are asymptotically independent in the upper tails i.e. the limit of $\chi^{(\bX)}(p)$ is zero.
We simulate $1000$ i.i.d. copies of $\bX$ from each model (a)-(c), with the sample size compared to the real data example, and we estimate   model \eqref{eq:biv} using the two-step procedure illustrated in Section \ref{sec:inference}. 
In particular, we use a cubic spline basis with basis dimension $K=10$ the default value in the \texttt{mgcv} R package.

\begin{figure}[h!]
	\begin{center}
		\begin{tabular}{cc}
			\includegraphics[width=0.45\linewidth]{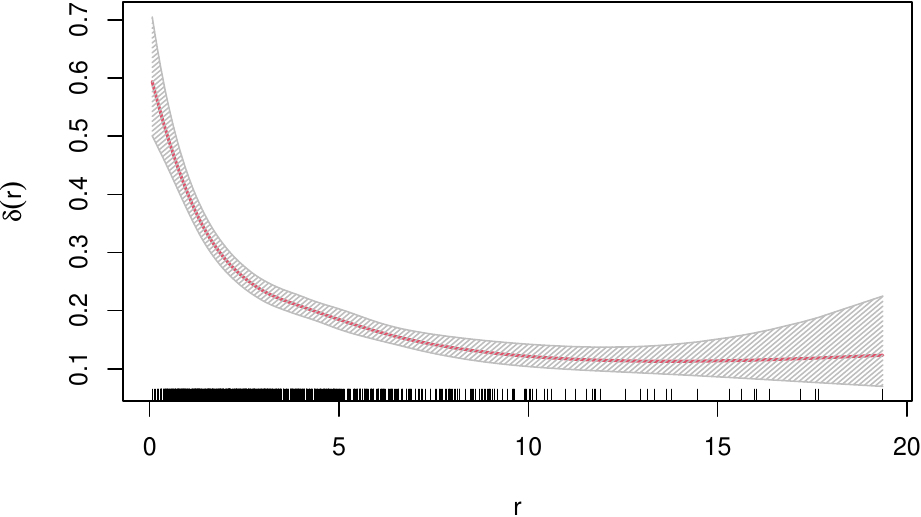} 
			&
			\includegraphics[width=0.45\linewidth]{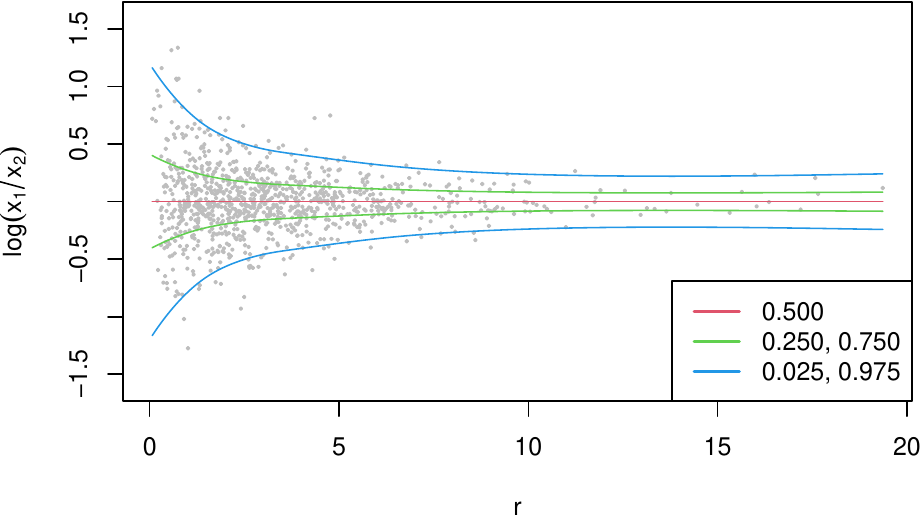}\\
			
			\includegraphics[width=0.45\linewidth]{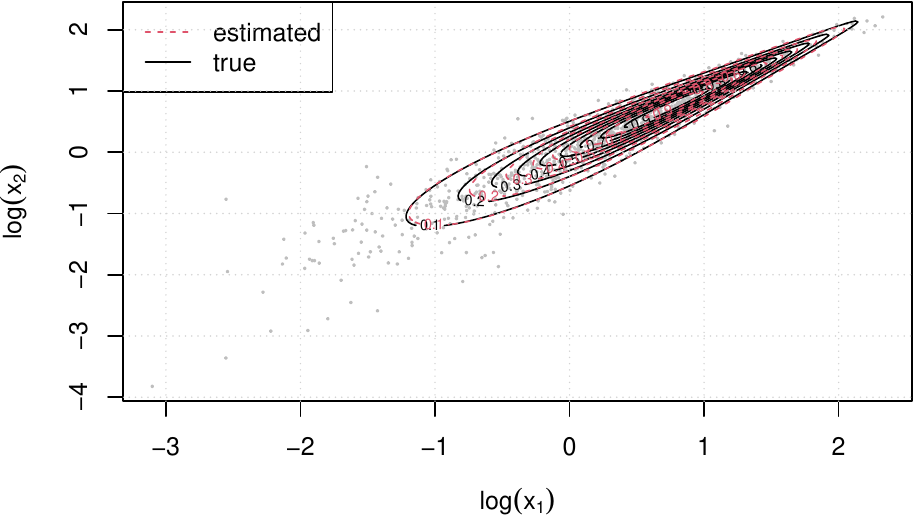} &
			\includegraphics[width=0.45\linewidth]{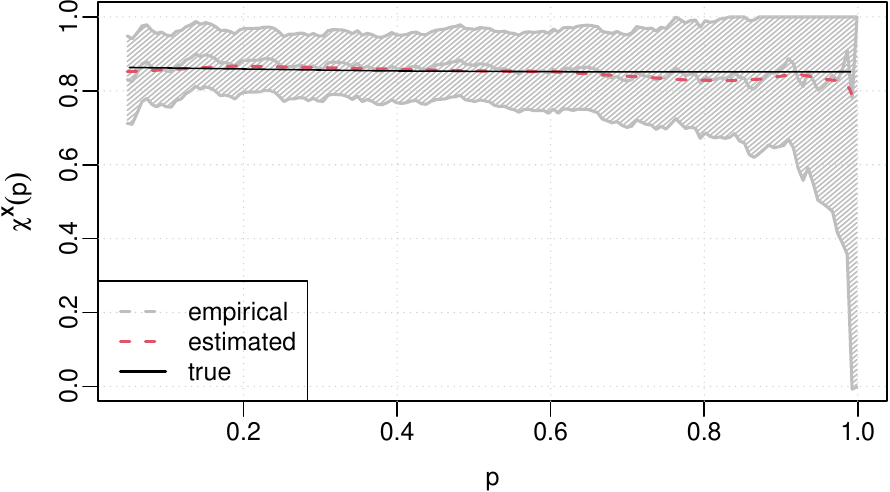} 
		\end{tabular}
		\caption{Bivariate symmetric logistic copula.
			Top left: the estimates of the standard deviation $\delta(r)$, $r=\norm{\bx}$,  with a shaded area that represents the 95\% pointwise confidence interval for the estimates.
			Top right: the scatterplot  of $\log(x_{1}/x_{2})$ versus $r$. The lines indicate the estimated quantile of the conditional Gaussian distribution.
			Bottom left: the contour plots of the true bivariate density and the estimated density. Bottom right: The true and estimated extremal dependence function $\chi^{(\bX)}(p)$.}
		\label{fig:logistic-example}
	\end{center}
\end{figure}

\begin{figure}[h!]
	\begin{center}
		\begin{tabular}{cc}
			\includegraphics[width=0.45\linewidth]{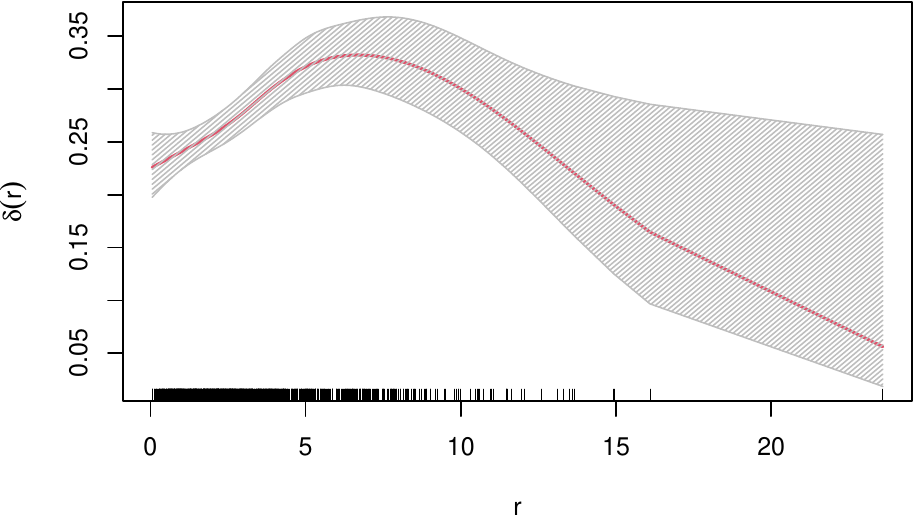} 
			&
			\includegraphics[width=0.45\linewidth]{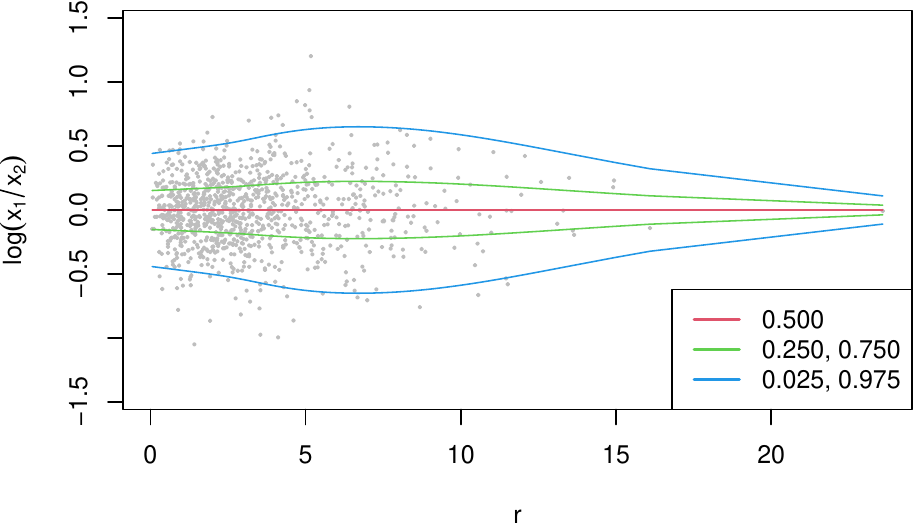}\\
			
			\includegraphics[width=0.45\linewidth]{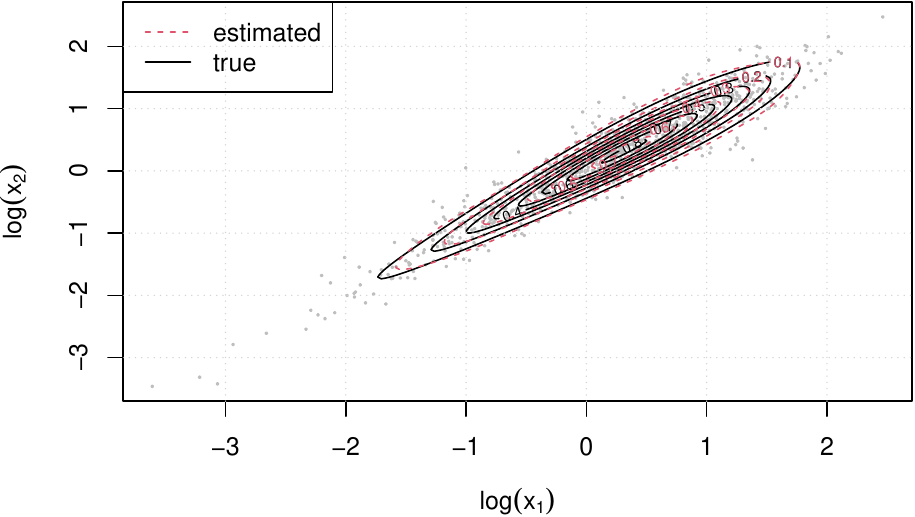} &
			\includegraphics[width=0.45\linewidth]{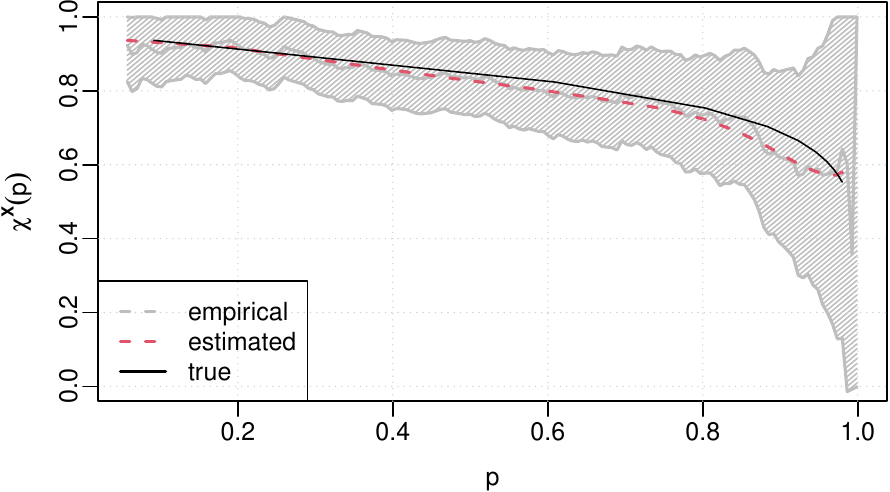} 
		\end{tabular}
		\caption{Inverted bivariate symmetric logistic copula.
			Top left: the estimates of the standard deviation $\delta(r)$,  $r=\norm{\bx}$  with a shaded area that represents the 95\% pointwise confidence interval for the estimates.
			Top right: the scatterplot  of $\log(x_i/x_2)$ versus $r$. The lines indicate the estimated quantile of the conditional Gaussian distribution.
			Bottom left: the contour plots of the true bivariate density and the estimated density. Bottom right: The true and estimated extremal dependence function $\chi^{(\bX)}(p)$.}\label{fig:invlogistic-example}
	\end{center}
\end{figure}

\begin{figure}[h!]
	\begin{center}
		\begin{tabular}{cc}
			\includegraphics[width=0.45\linewidth]{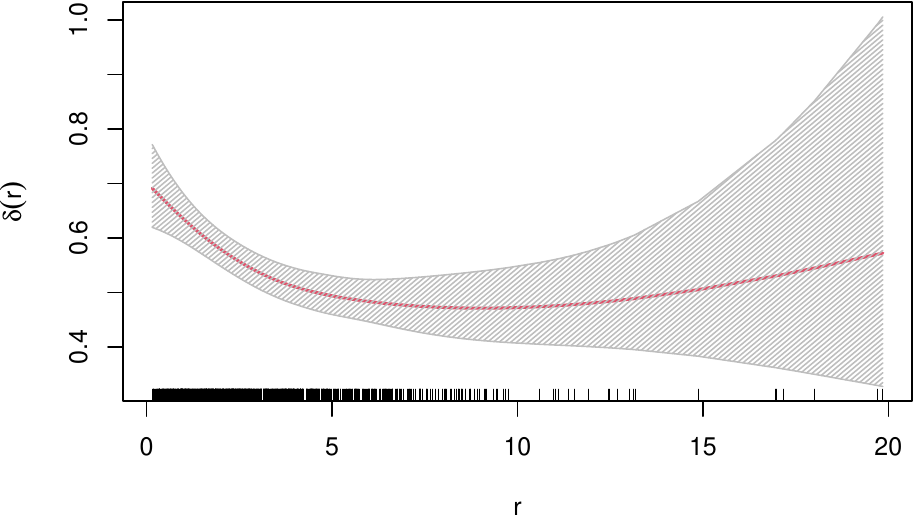} 
			&
			\includegraphics[width=0.45\linewidth]{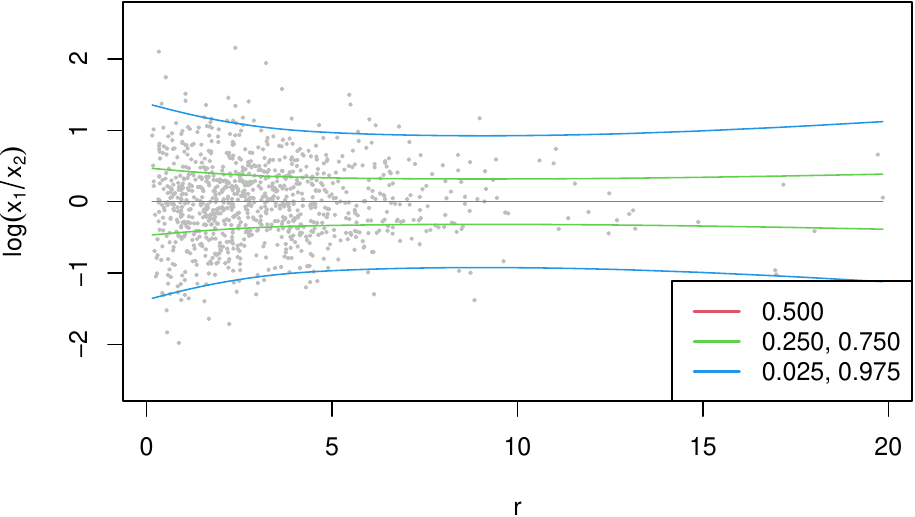}\\
			
			\includegraphics[width=0.45\linewidth]{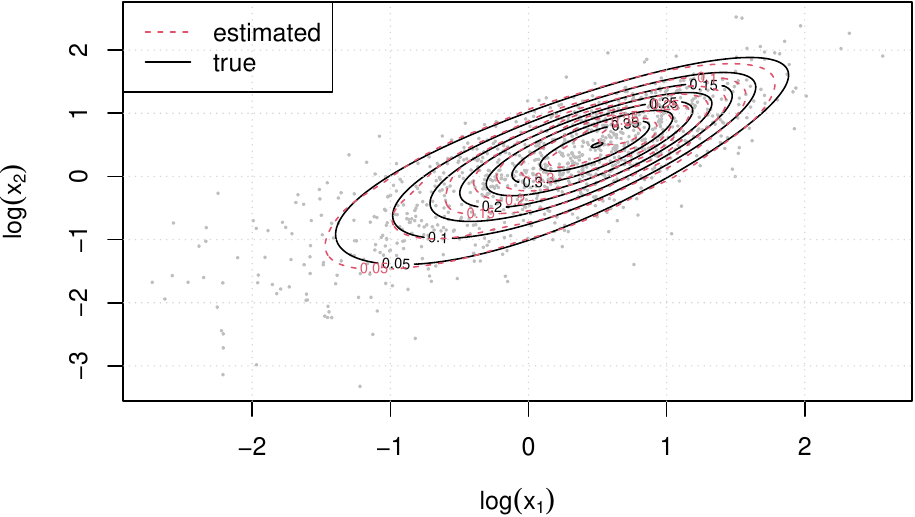} &
			\includegraphics[width=0.45\linewidth]{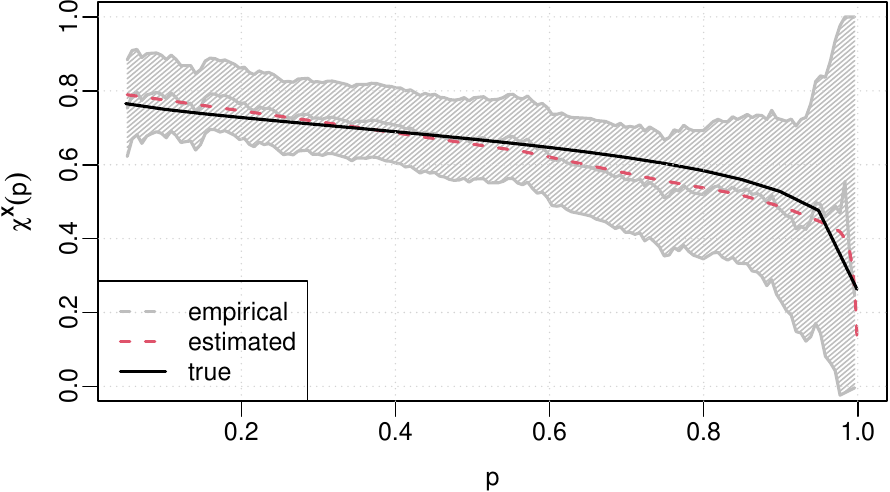} 
		\end{tabular}
		\caption{Gaussian copula.
			Top left: the estimates of the standard deviation $\delta(r)$, $r=\norm{\bx}$,  with a shaded area that represents the 95\% pointwise confidence interval for the estimates.
			Top right: the scatterplot  of $\log(x_1/x_2)$ versus $r$. The lines indicate the estimated quantile of the conditional Gaussian distribution.
			Bottom left: the contour plots of the true bivariate density and the estimated density. Bottom right: The true and estimated extremal dependence function $\chi^{(\bX)}(p)$.}\label{fig:gaussian-example}
	\end{center}
\end{figure}

For each example, a set of figures is provided in the text (see 
Figures \ref{fig:logistic-example},  \ref{fig:invlogistic-example}, and  \ref{fig:gaussian-example}). These figures report the estimate of the standard deviation, a scatterplot of the values of $\log(x_i/x_d)$ versus $r_i$ together with the estimated quantile of the conditional Gaussian distribution, denoted by $\mathcal{N}( 0,\delta(r)^2)$, and a comparison between the true bivariate density and the estimated one. Finally, a plot of the true and estimated extremal dependence function $\chi^{(\bX)}(p)$ is provided 
It must be acknowledged that this experiment is limited in scope. However, it can be observed that the model exhibits sufficient flexibility to accommodate the three scenarios, particularly in the context of asymptotic dependence.

\subsection{UK River Wye  data}
To illustrate our method, we analyse river discharge recordings at three locations on the same river network in Great Britain.
The UK National River Flow Archive (NRFA) collects data from over 1,500 stations that measure   water flows in rivers, lakes, and reservoirs (see \texttt{https://nrfa.ceh.ac.uk/}). Three stations,  Erwood, Redbrook and Ddol Farm  along the River Wye,  are of particular interest because large  flooding events can occur in this hydrological basin. 

\begin{figure}[h!]
	\begin{center}
		\begin{tabular}{cc}		
			\includegraphics[width=0.35\linewidth,height=0.5\textwidth]{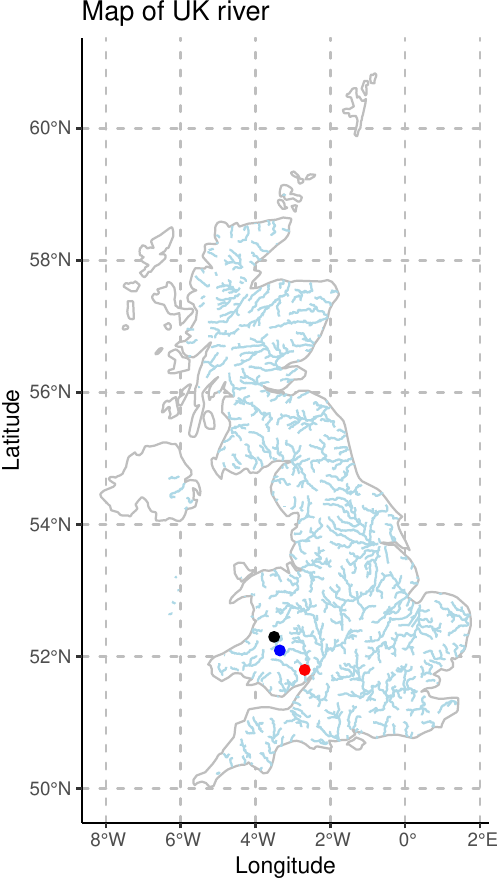}
			&	
			\includegraphics[width=0.6\linewidth,height=0.5\textwidth]{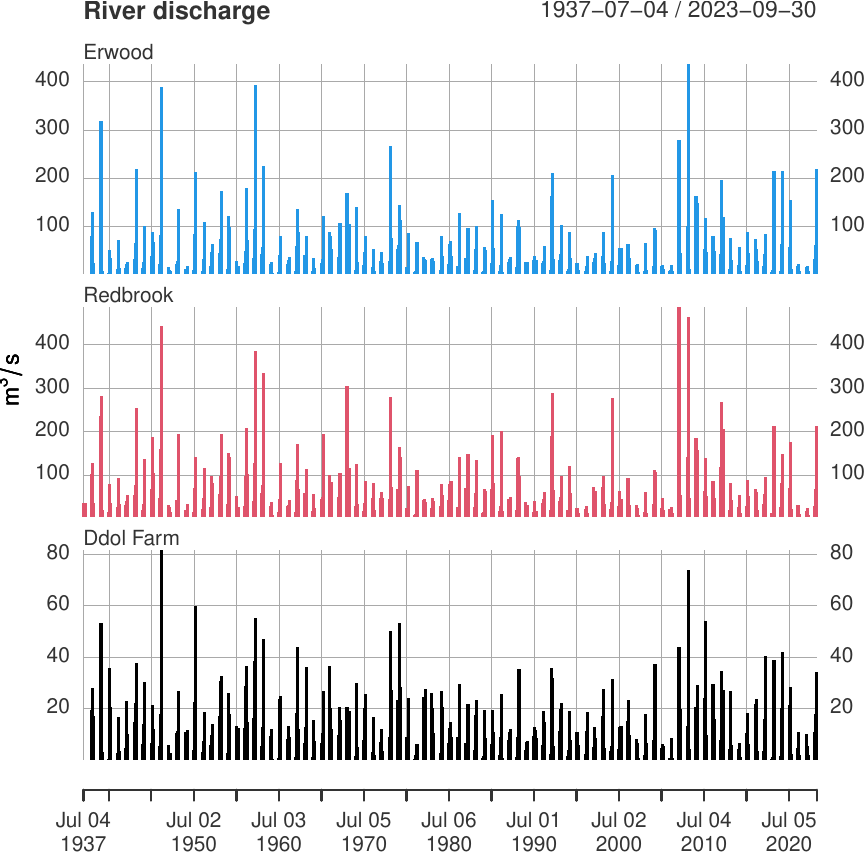}
		\end{tabular}
		\caption{Right: the UK National River Flow map with Erwood (blue), Redbrook  (red) and Ddol Farm (black) stations at Wye river. Left: Time series of weekly maximum summer discharges.}\label{fig:wye-tri}
	\end{center}
\end{figure}

In order to eliminate the seasonal effect and reduce temporal clustering, the focus is directed towards weekly maximum summer (June-August) discharges from July 1938 to August 2023.    This constitutes a sample of 1\,131 observations for each station, as illustrated in Figure \ref{fig:wye-tri}.
It is evident that the three nearby recordings are indeed dependent,

For each of the stations, each time series is re-scaled by dividing by its empirical median. Subsequently, the minimum values recorded at each station were subtracted from the observations.

After this location scale adjustment, we identify the transformed value of Erwood, Redbrook, and Ddol Farm stations as $X_1$, $X_2$, and $X_3$, respectively and we consider the transformed data $(r_i, v_{i1},v_{i2})=(\|x_i\|,\log(x_{i,1}/x_{i,3}), \log(x_{i,2}/x_{i,3})))$, $i=1,\ldots, 1131$.

We fit a trivariate  EGPD as in Section \ref{sec:inference} by setting  
the number of Bernstein basis functions to $m=\lfloor 0.5 n/\log(n)\rfloor=80$, in \eqref{eq: F bar} \citep{Naveau25}, and the number of cubic spline basis functions to $K=12$,  in \eqref{eq:link} .
\begin{table}[h]
	\begin{center}
		\begin{tabular}{|c|c|c|}
			\hline
			$\kappa$ & $\xi$ &$\rho$\\ 
			\hline
			2.23 & 0.37 & 0.67\\
			(0.32, 3.91) & (0.21, 0.41)& (0.64,0.71)\\
			\hline
		\end{tabular}	
	\end{center}	
	\caption{Estimated values of the parameters in \eqref{eq:gaussian_pdf_logistic_model}. The values between parentheses denote a  $0.95$ bootstrap pivotal confidence interval obtained using a parametric bootstrap with a sample size of 1\,000.}
\end{table}	

\begin{figure}[h!]  
	\begin{center}
		\begin{tabular}{cc}
			$\norm{\bx}$ & $1/\norm{\bx}$\\
			\includegraphics[width=0.45\linewidth,height=0.2\textheight]{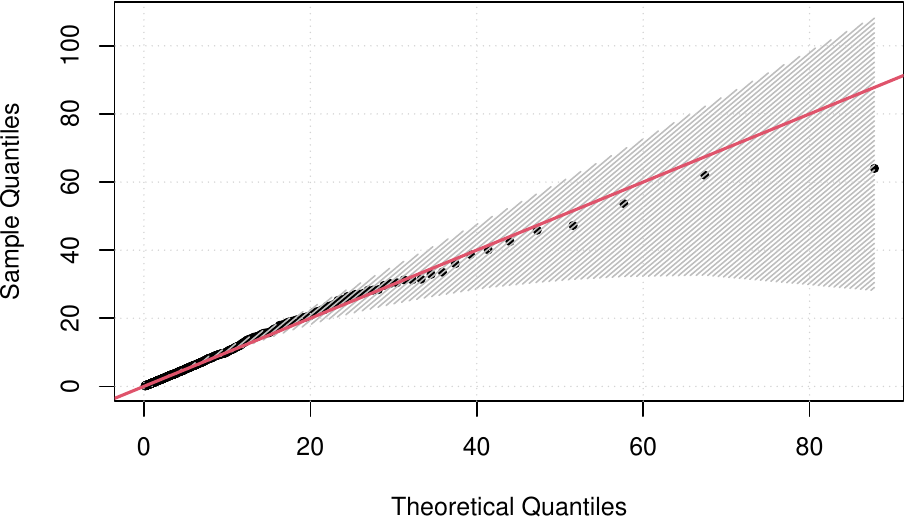} &			\includegraphics[width=0.45\linewidth,height=0.2\textheight]{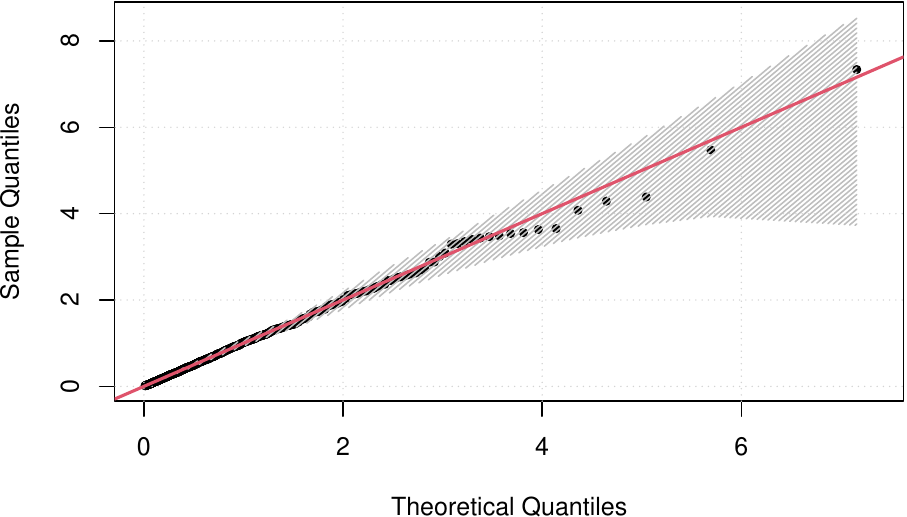}
		\end{tabular}
	\end{center}
	\caption{Left: plot of the  empirical quantiles of $\|\bx\|$ against the estimated quantile. Right: plot of the empirical quantiles of $1/\norm{\bx}$ against the estimated quantile.
		The filled region represents pointwise 0.95 bootstrap pivotal confidence interval obtained using a parametric bootstrap with a sample size of $1000$.
	}\label{fig:wye-r-tri}
\end{figure}

The estimates of $\kappa$ and $ \xi$ provide a high degree of agreement with the data at hand.  Specifically, the quantile-quantile plots in Figure \ref{fig:wye-r-tri} indicate that our EGPD  model for the radius captures the distribution of $R$   across the entire value range, including the central tendency and extremes.
This is especially true for the plot of the quantiles of $1/r_i$.

\begin{figure}[h!]  
	\begin{center}
		\begin{tabular}{cc}
			{\includegraphics[width=0.45\linewidth,height=0.2\textheight]{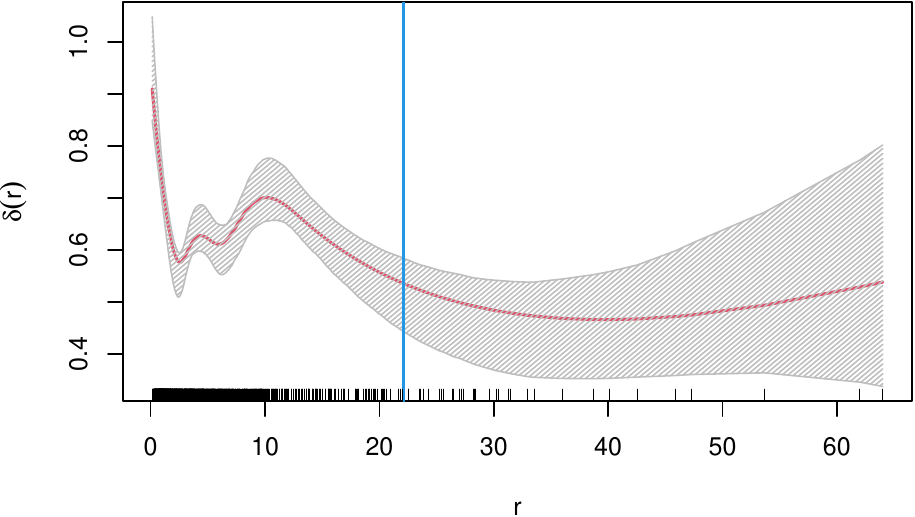}} &
			\includegraphics[width=0.45\linewidth,height=0.2\textheight]{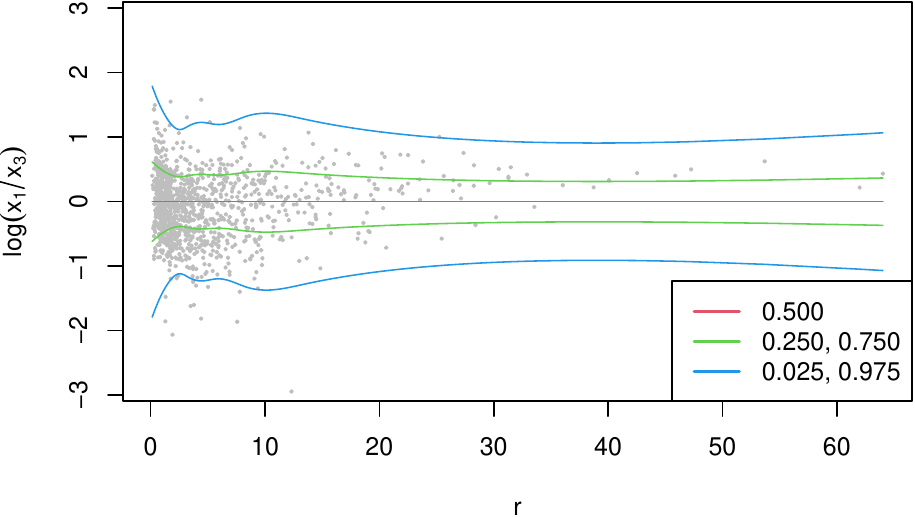} \\
			\hspace{0.8cm} (a) & \hspace{0.8cm} (b) \\
			\includegraphics[width=0.45\linewidth,height=0.2\textheight]{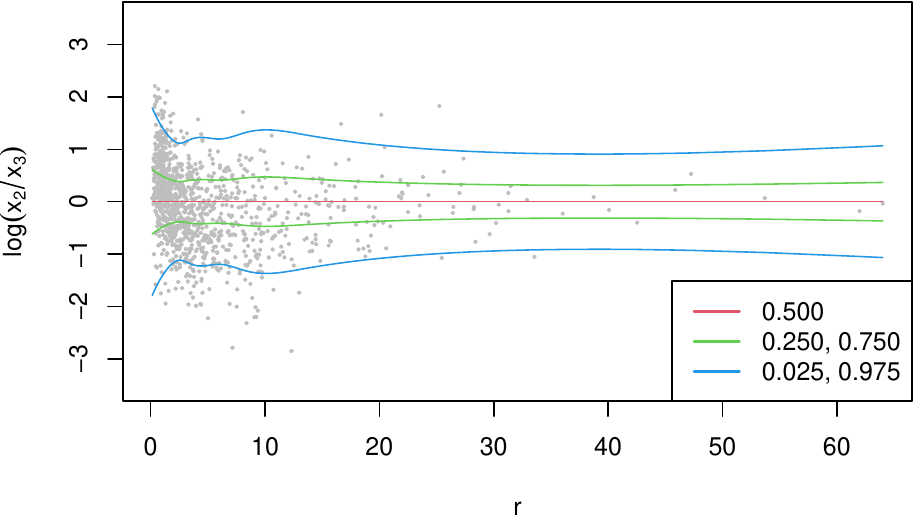} &
			\includegraphics[width=0.45\linewidth,height=0.2\textheight]{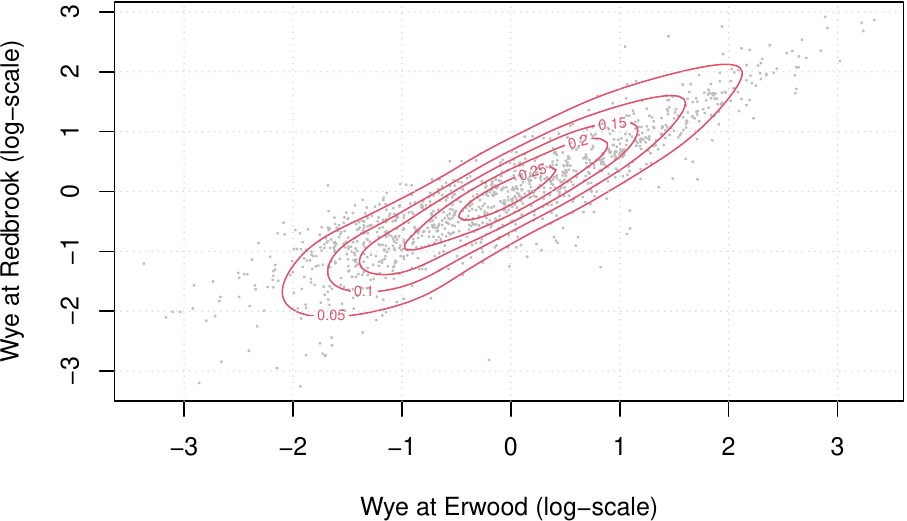} \\
			\hspace{0.8cm} (c) & \hspace{0.8cm} (d) \\
			\includegraphics[width=0.45\linewidth,height=0.2\textheight]{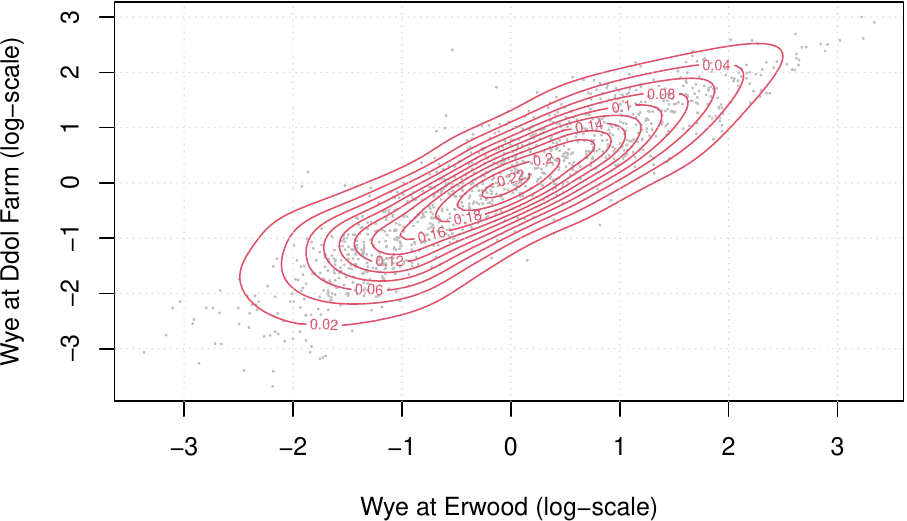} &
			\includegraphics[width=0.45\linewidth,height=0.2\textheight]{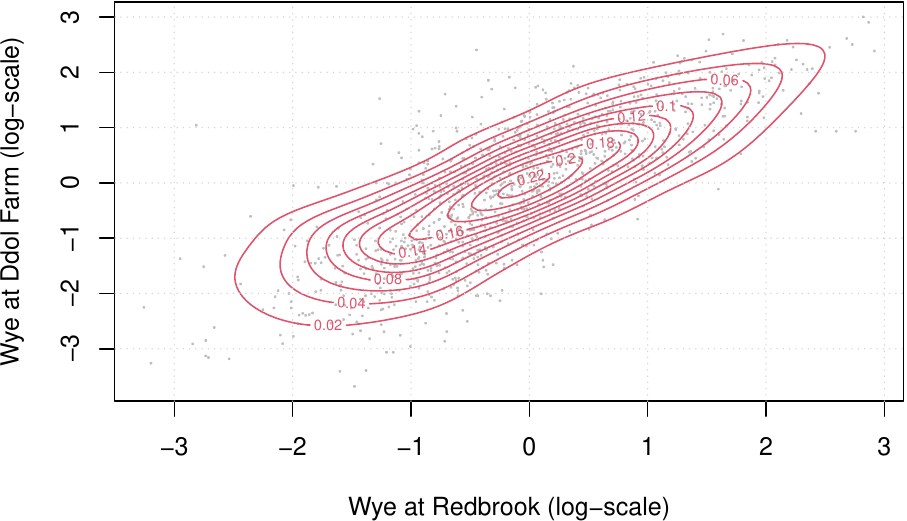}\\
			\hspace{0.8cm} (e) & \hspace{0.8cm} (f) \\
		\end{tabular}
	\end{center}
	\caption{(a) The estimated conditional  standard deviation $\delta(r)$ with $r=\norm{\bx}$. The filled region represents pointwise $0.95$ bootstrap pivotal confidence interval obtained using a parametric bootstrap with a sample size of $1000$. 
		The blue vertical line indicates the 0.97 empirical quantile of the $r$ distribution. (b)-(c) Scatterplot of the transformed angular components $\log(x_{1}/x_{3}),\log(x_{2}/x_{3})$ against the radius $r$.
		The solid lines represent different quantiles of the estimated Gaussian distribution of $[\log(X_i/X_d)|R=r]$.
		(d)-(e)-(f) Estimated bivariate densities for each pair of stations. }\label{fig:wye-v-tri}
\end{figure}

In terms of dependence, the diagnostic plots in Figure \ref{fig:wye-v-tri} show a non-monotonic pattern in the estimate of the conditional standard deviation of $V|R=r$, $\hat{\delta}(r)$ as the value of $r$ increases.
However, when considering the 95\% confidence intervals, the graph appears to be constant beyond the 0.97 quantile of the empirical distribution of $r_i$. 

The scatter plots of $\log(x_1/x_3)$ (Figure \ref{fig:wye-v-tri}-(b)) and $\log(x_2/x_3)$ (Figure \ref{fig:wye-v-tri}-(c)) versus $r_i$ show that the distribution around zero is moderately asymmetric.  Conversely, the estimated bivariate densities $(X_i,X_j)$, $i\ne j$, demonstrate a high degree of agreement with the data. It is noteworthy that the density is represented on a logarithmic scale to more effectively capture the pattern.

\section{Conclusions}\label{sec:conclusions}
In general, there is no unique solution to extend a univariate distribution to the multivariate context. This assertion is also applicable to the EGPD.

The present proposal breaks down the multivariate data into radial and angular components. The radial component is modelled using a univariate EGPD, while the angular distribution varies according to specific conditions. Following this decomposition, a variety of options for the angular component's distribution become available.

In practice, we model log-ratios with for a logistic-heteroscedastic  structure  based on a classical multivariate Gaussian, see \eqref{eq:gaussian_pdf_logistic_model}. 
This has the advantage to keep a simple and fast inference scheme, while offering flexible multivarariate pdfs.  
Moreover the Gaussian distribution makes it possible to use existing R software functions to estimate parameters, thus simplifying the coding task. 
Finally, the potential exists for the implementation of vector-valued Gaussian processes as a means of modelling serial \citep{lutkepohl2005} or spatial dependence  \citep{gelfand2010multivariate} in the data.

To make the construction proposed in Definition \ref{def: BEGPD} even more flexible, a mixture of Dirichlet distributions for the angular component could be implemented, as any angular distribution can be approached arbitrarily well by a mixture of Dirichlet distributions, as shown in \citet{sabourin:naveau:2014}, for instance.

It is worth to mention that other alternatives exists. 
For example, \citet{Alotaibi-et-al:2025} recently suggested a different approach. 
Contrary to our present work, they modeled the vector of interest as a random  sum of two components: 
one for the lower tail and another for the upper tail.  The contribution of each component depends on the radius. In addition, the inference scheme differs as they do not access to the likelihood function. 
Consequently, they estimated their model's parameters using generative    methods based on  amortized neural inference
\cite{Zammit-Mangion.etal:2025}. 
Finally, a comprehensive review comparing   all recent attempts sharing the same goal of avoiding threshold selections to model the full multivariate distribution will be welcome. But, at this stage, it is  beyond the scope of this article.  

\section*{Acknowledgments}
Part of Naveau’s research work was supported by European H2020 XAIDA (Grant agreement ID: 101003469) and the  French  Agence Nationale de la Recherche:  EXSTA, the PEPR TRACCS   (PC4 EXTENDING, ANR-22-EXTR-0005),  the PEPR IRIMONT (France 2030 ANR-22-EXIR-0003) SHARE, the  PEPR Maths-Vives (ANR-24-EXMA-0008).

Both authors have also received support from the Geolearning research chair, a joint initiative of Mines Paris and the French National Institute for Agricultural Research (INRAE).

Finally, we thank Prof. Philippe Soulier for his valuable explanations concerning the upper tail behavior of the harmonic mean under hidden regular variations.

\section*{Appendix}

\paragraph{Proof of Lemma \ref{lem:1/X}:} 
Let $\tilde{F}$ denote the cdf of $1/X$. It is always possible to define the function 
$$
\tilde{B}(u):=\tilde{F}\left( H^{-1}_{1/\kappa}\left( u^{\xi} \right) \right).
$$
As a composition of non-decreasing functions, $\tilde{B}(u)$ is also non-decreasing for $u$ on $[0,1]$ and, by definition, we have
as $\tilde{F}(x)= \tilde{B}\left(H_{1/\kappa}^{1/\xi}(x)\right)=1-F\left({1}/{x}\right)$.  
This gives that 
$$
\lim_{x \to 0} \frac{\tilde{F}(x)}{x^{1/\xi}}= \lim_{x \to \infty} \frac{\overline{F}(x)}{x^{-1/\xi}}= \kappa \; \xi^{1/\xi} \; b(1)
\mbox{ and }
\lim_{x \to \infty} \frac{\overline{\tilde F}(x)}{1/\xi \overline{H}_{1/\kappa}(x)}= 
\lim_{x \to  0} \frac{F(x)}{1/\xi \overline{H}_{1/\kappa}(1/x)}= \xi \; b(0).
$$
To show \eqref{eq: b to b tilde}, we write 
$$
\tilde{f}(x)= 1/\xi \cdot h_{1/\kappa}(x)     \cdot H^{1/\xi-1}_{1/\kappa}(x) \cdot \Tilde{b}\left( H^{1/\xi}_\xi(x)\right), \mbox{ for any} x \geq  0.
$$
It follows that 
$$
\lim_{x \to 0} \frac{\tilde{F}(x)}{x^{1/\xi}}=\lim_{x \to 0}\frac{\tilde{f}(x)}{1/\xi x^{1/\xi-1}}=    \tilde{b}(0).
$$
and 
$$
\lim_{x \to \infty} \frac{\overline{\tilde F}(x)}{1/\xi \; \overline{H}_{1/\kappa}(x)}=
\lim_{x \to \infty} \frac{\tilde{f}(x)}{1/\xi \; h_{1/\kappa}(x)}= \tilde{b}(1).
$$
The expected result follows.

\hfill $\square$ 

\bigskip
\paragraph{Proof of Lemma \ref{lem: norm(X) EGPD}:}

Let $F_d$ denote the cdf of $\Rad$. It is always possible to define the function 
$$
B_d(u)=F_d\left( H^{-1}_{\xi}\left( u^{1/\kappa_d} \right) \right)
\mbox{ with } \kappa_d =a \; \kappa.
$$
As a composition of non-decreasing functions, $B_d(u)$ is also non-decreasing for $u$ on $[0,1]$ and, by definition, we have
as $F_d(x)= B_d\left(H_{\xi_d}^{\kappa_d}(x)\right)$.  
From \eqref{eq: b to b tilde}, we can then write 
$$
\lim_{x \to \infty} \frac{\Pr(\Rad >  x)}{\kappa_d  \overline{H}_{\xi}(x)}=
c_+  \lim_{x \to \infty} \frac{\overline{F}(x)}{\kappa_d \overline{H}_{\xi}(x)} = 
c_+  \frac{1}{a}b(1)= b_d(1).
$$
Concerning the lower tail, we have 
\begin{eqnarray*}
	b_d(0)&=& \lim_{x \to  0} \frac{\Pr( \Rad \leq  x) }{x^{\kappa_d}}, \mbox{ by definition of $b_d(0)$},\\
	&=&c_-  \lim_{x \to  0} \frac{[\Pr( X_i \leq  x)]^a }{x^{\kappa_d}}, \mbox{ by \eqref{eq: tail equivalence conditions lower}},\\     
	&=&c_-  \lim_{x \to  0} \frac{\left( b(0) x^{\kappa}\right)^{a} }{x^{\kappa_d}}, \mbox{ as $X_i$ follows a MGPD($\kappa,B,\xi)$},\\
	&=&  c_-\; b^a(0), \mbox{ when $\kappa_d= a \; \kappa$.} 
\end{eqnarray*}
This completes the proofs for $\Rad$.
Concerning $|\bY|$, we just need to apply Lemma \ref{lem:1/X}.

\hfill $\square$

\paragraph{Proof of Proposition \ref{prop:convergence}}:
Here, we note $v_i=\log(x_i/x_d)$.
The joint pdf of \((R, V_1, \ldots, V_{d-1})\T\) can be denoted as 
$$g(r, \bv):= f_{R,V_1,\ldots,V_{d-1}}(r, v_1, \ldots, v_{d-1}),$$
and the pdf of $\bX$ can be expressed as  
\[
f(\bx) = g(r(\bx), \bv(\bx)) \cdot \left| \det (J) \right|
\]
with
\[
\begin{aligned}
	r(\bx) &= x_1 + \cdots + x_d \\
	v_i(\bx) &=\log{x_i}-\log{x_d}, \quad i=1,\ldots, d-1,
\end{aligned}
\]

$$
\bJ=
\left(
\begin{array}{ccccc}
	1 & 1& \ldots & 1&1\\
	1/{x_1} & 0& \ldots & 0&-1/x_d\\
	0& 1/{x_2} &  \ldots & 0&-1/x_d\\
	\vdots & \vdots &\ddots &\vdots &\vdots \\
	0 & 0& \ldots & 1/x_{d-1}&-1/x_d\\
\end{array}
\right),
$$

and 
\[
\left| \det (\bJ) \right| = \frac{\sum_{i=1}^{d}x_i}{\prod_{i=1}^{d}x_i}
\]
Hence, 
\[
f(\bx) = g(r, \log(x_1/x_d), \ldots, \log(x_{d-1}/x_d))\, \frac{\sum_{i=1}^{d}x_i}{\prod_{i=1}^{d}x_i}
\]
As  $V_i = \delta(r) Z_i$, 
$$
g(r, \bv) = f_{{\bV} \text{given } \norm{\bX}=r}(\bv)   f_{\norm{\bX}}(r)= \frac{1}{\delta^{d-1}(r)} f_{\bZ}\left(\frac{\log(x_1/x_d), \ldots, \log(x_{d-1}/x_d)}{\delta(r)}\right) f_{\norm{\bX}}(r),
$$
the expression  \eqref{eq:pdf_logistic_model}   follows. 

We also have,  with the notation $V(t)=t^{-1/\xi}$, a regularly varying function of order $-1/\xi$, that 

\[
\frac{f(t \bx)}{t^{-d} V(t)}
=  \frac{\sum_{i=1}^{d}x_i}{\prod_{i=1}^{d}x_i}
\frac{1}{\delta^{d-1}(t \; r)} f_{\bZ}\left(\frac{\log(x_1/x_d), \ldots, \log(x_{d-1}/x_d)}{\delta(t \; r)}\right) \left( \frac{f_{\norm{\bX}}(t\;r)}{t^{-1} V(t)} \right),
\]
As $\norm{\bX}$ follows a EGPD, it satisfies \eqref{eq: b to b tilde} and 
$$
\frac{f_{\norm{\bX}}(t\;r)}{t^{-1} V(t)}  
\sim b(1)\xi^{-1/\xi-1} \frac{(t \; r)^{-1/\xi-1}}{t^{-1/\xi-1}}= b(1)\xi^{-1/\xi-1} \norm{\bx}^{-1/\xi-1}.
$$
Hence, we see the link with \eqref{eq: pdf + } and we notice that for any $\bx \in [\bzero,\infty) \setminus \{\bzero\}$ and $t>0$,
$$
\lambda_+(t \bx) = t^{-1/\xi-d} \lambda_+(\bx). 
$$
Condition \eqref{eq: pdf + } corresponds to Equation (6.35), with $V(t)=t^{-1/\xi}$,  from 
Theorem 6.4  of \cite{Resnick:2007}, i.e. the form 
$$
\lim_{t \rightarrow \infty} \sup_{\norm{\bx}=1} \Bigg|   
\frac{f(t \bx)}{t^{-d} V(t)} -\lambda_+(\bx)  \Bigg|=0.
$$

\hfill $\square$

\bibliographystyle{abbrvnat}
\bibliography{biblio}

\end{document}